\documentclass[12pt]{article}

\usepackage[a4paper,left=30mm,right=20mm,top=20mm,bottom=20mm]{geometry}
\usepackage{graphics}
\usepackage{amsmath}
\usepackage{epsfig}
\usepackage{cite}
\usepackage{xcolor}

\newcommand{\bea}{\begin{eqnarray}}
\newcommand{\eea}{\end{eqnarray}}
\newcommand{\be}{\begin{equation}}

\newcommand{\ee}{\end{equation}}
\newcommand{\MSbar}{\overline{\rm MS}}

\title{Transverse momentum dependent parton densities in a proton from the generalized DAS approach}

\author{A.V.~Kotikov$^{1,2}$, A.V.~Lipatov$^{2,3}$, B.G.~Shaikhatdenov$^{2}$, P.~Zhang$^{4}$ }

\begin{document}

\maketitle

\begin{center}
{\it $^{1}$Institute of Modern Physics, Lanzhou 730000, China}\\
{\it $^{2}$Joint Institute for Nuclear Research, 141980, Dubna, Moscow region, Russia}\\
{\it $^{3}$Skobeltsyn Institute of Nuclear Physics, Lomonosov Moscow State University, 119991, Moscow, Russia}\\
{\it $^{4}$ School of Physics and Astronomy, Sun Yat-sen University, Zhuhai 519082, China}

\end{center}

\vspace{0.5cm}

\begin{center}

{\bf Abstract }

\end{center}

\indent
We use the Bessel-inspired behavior of parton densities at small Bjorken $x$ values,
obtained in the case of the flat initial conditions for DGLAP evolution equations
in the
double scaling QCD approximation (DAS), to evaluate the transverse momentum dependent (TMD, or unintegrated) quark and gluon
distribution functions in a proton. The calculations are performed analytically using
the Kimber-Martin-Ryskin (KMR) prescription with different implementation of
kinematical constraint, reflecting the angular and strong ordering conditions.
The relations between the differential and integral formulation of the KMR approach
is discussed. Several phenomenological applications of the proposed TMD parton densities
to the LHC processes are given.

\vspace{1.0cm}


\newpage

\section{Introduction} \indent

A theoretical description of a number of high energy processes at hadron colliders which
proceed with large momentum transfer and containing multiple hard scales can be obtained with
transverse momentum  dependent  (TMD), or unintegrated, parton (quark and/or gluon) density functions in a proton\cite{1}.
These quantities depend on the fraction $x$ of the proton longitudinal momentum
carried by a parton, the two-dimensional parton transverse momentum ${\mathbf k_T^2}$ and
hard scale $\mu^2$ of the hard process and encode non-perturbative information on proton structure,
including transverse momentum and polarization degrees of freedom.
They are related to the physical cross sections and
other observables,
measured in the collider experiments, via TMD factorization theorems in Quantum Chromodynamics (QCD).
The latter provide the necessary framework to separate hard partonic physics, described with a
perturbative QCD expansion, from soft hadronic physics.
At present, there are number of factorization approaches which incorporate the transverse momentum dependence in the
parton distributions: for example, the Collins-Soper-Sterman approach\cite{2} (or TMD factorization), designed
for semi-inclusive processes with
a finite and non-zero ratio between the hard scale $\mu^2$ and total energy $s$, and
high-energy factorization\cite{3} (or $k_T$-factorization\cite{4}) approach, valid in
the limit of a fixed hard scale and high energy.

In the high-energy factorization, the TMD gluon density satisfies the
Balitsky-Fadin-Kuraev-Lipatov (BFKL)\cite{5}
or Ciafaloni-Catani-Fiorani-Marchesini (CCFM)\cite{6} evolution equations,
which resum large terms
proportional to $\alpha_s^n \ln^n s \sim \alpha_s^n \ln^n 1/x$, important at
high energies $s$ (or, equivalently, at small $x$).
Thus, one can effectively take into account higher-order radiative corrections
to the production cross sections (namely, a part
of NLO + NNLO +... terms corresponding to the initial-state real gluon
emissions).
The CCFM equation
takes into account additional terms
proportional to $\alpha_s^n \ln^n 1/(1 - x)$ and therefore
is valid at both small and large $x$.
The number of phenomenological applications of the high-energy factorization
and CCFM evolution
is known in the literature (see, for example,\cite{7,8,9,10,11,12,13,14,15,16,17} and references therein).

There are also other approaches determining the TMD gluon and quark density functions in a proton.
So, one can evaluate them
using the schemes based on the
conventional Dokshitzer-Gribov-Lipatov-Altarelli-Parisi (DGLAP)\cite{18}
equations, namely the Parton Branching (PB) approach\cite{19,20} and
Kimber-Martin-Ryskin (KMR) prescription\cite{21}.
Former gives numerical iterative
solution of the DGLAP evolution equations for collinear and TMD parton density functions upon using a
concept of resolvable and non-resolvable branchings and by applying Sudakov formalism
to describe the parton evolution from one scale to
another without resolvable branching.
The
splitting kinematics at each branching vertex
is described by the
DGLAP equations and angular ordering condition for parton emissions
can be applied instead of usual DGLAP ordering in virtuality.
The latter is a formalism invented for constructing
the TMD parton distributions from well-known conventional (collinear) parton density functions (PDFs)
under the key assumption that the
transverse momentum dependence of the parton distributions enters only at the
last evolution step.
The KMR procedure is believed to take into account effectively
the major part of next-to-leading logarithmic (NLL) terms $\alpha_s (\alpha_s \ln\mu^2)^{n-1}$
compared to the leading logarithmic approximation (LLA), where
terms proportional to $\alpha_s^n \ln^n \mu^2$ are taken into account.
The KMR approach is currently explored\cite{22} at
next-to-leading order (NLO) and commonly used in the phenomenological applications
(see, for example,\cite{8,10,11,12,13,14,15,16} and references therein),
where the standard proton PDFs (as obtained, for example,
by the NNPDF\cite{23} or CTEQ\cite{24} Collaborations) were
taken as an input numerically.
The relation between the PB and KMR scenarios was
discussed \cite{25} and the connection between the PB and CCFM approaches
was established very recently\cite{26}.

The  KMR formalism is used in the present paper for analytical calculations
of the TMD quark and gluon distributions in a proton.
The calculations are based on the expressions\cite{Cvetic1,Q2evo,HT}
for conventional PDFs
obtained in the {\it generalized} double asymptotic scaling (DAS) approximation\cite{Munich,Q2evo,HT}.
The latter is connected to the asymptotic behaviour of the DGLAP evolution discovered many
years ago\cite{Rujula}. As it was shown,
flat initial conditions for DGLAP equations, applied in the {\it generalized} DAS scheme,
lead to the Bessel-like behaviour for the proton PDFs at small $x$.
Using the results\cite{Cvetic1,Q2evo,HT}, we derive the analytical expressions for the TMD
quark and gluon densities at leading order (LO) and present them in a quite compact form.
We implement different treatment
of the kinematical constraint involved in the KMR prescription
(namely, angular and strong ordering conditions) and
discuss the relations between the differential and integral formulation
of the KMR procedure pointed out recently in\cite{Golec-Biernat:2018hqo}.
Finally, we present some phenomenological applications of obtained TMD
parton densities to hard LHC processes, sensitive to the quark and gluon content of the proton.
To be precise, using the $k_T$-factorization QCD approach,
we consider the inclusive production of $b$-jets and $b \bar b$-dijets at $\sqrt s = 7$~TeV,
inclusive production of the Higgs bosons (in the diphoton decay mode) at $\sqrt s = 13$~TeV
and charm and beauty contributions to the deep inelastic proton
structure function $F_2(x,Q^2)$ at different values of $Q^2$.

The outline of our paper is following. In Section~2 we briefly describe
our theoretical input. The calculations are explained in detail
in Section~3,
where we also present the modifications of PDFs and TMDs at low $Q^2$-values and
outside of the standard small $x$ range.
Section~4 present
numerical results and discussions. Section~5 contains our conclusions.

\section{Theoretical framework} \indent

Since the KMR approach is based on the standard PDFs, here we present a review of small $x$ behaviour
of parton densities. Moreover, we introduce the basic formulas of the KMR approach itself.

\subsection{
      PDFs and proton structure function $F_2(x,Q^2)$
} \indent

The fairly
reasonable agreement between HERA data
\cite{33,34,35,36,37}
and the results of NLO perturbative QCD evaluations is observed for
$Q^2 \geq 2$ GeV$^2$ (see reviews\cite{CoDeRo,39} and references therein).
Therefore, it can be concluded that pQCD is capable of describing the evolution
of proton structure function $F_2(x,Q^2)$ and its derivatives
down to very low $Q^2$ values.

It was pointed out\cite{BF1} that the HERA small-$x$ data can be
well interpreted in
terms of the so-called doubled asymptotic scaling (DAS) phenomenon
related to the asymptotic
behaviour of the DGLAP evolution discovered many years ago\cite{Rujula}.
The study\cite{BF1} was extended\cite{Munich,Q2evo,HT}
to include the finite parts of anomalous dimensions (ADs)
of Wilson operators and Wilson coefficients\footnote{
In the standard DAS approximation\cite{Rujula} only the AD singular
parts
were used.}.
This led to predictions\cite{Cvetic1,Q2evo,HT} of the small-$x$ asymptotic
form of PDFs
in the framework of the DGLAP dynamics,
which were obtained
starting at some $Q^2_0$ with
the flat function
 \begin{equation}
f_a (x, Q^2_0) = A_a,
\label{1}
 \end{equation}
where $f_a$ are PDFs
multiplied by $x$, $a = q$ or $g$
and $A_a$ are unknown parameters to be determined from the data.
We refer to the approach of\cite{Munich,Q2evo,HT}
as {\it generalized} DAS approximation. In this approach
the flat initial conditions~(\ref{1}) determine the
basic role of the AD singular parts
as in the standard DAS case,
whereas the contributions coming from AD finite parts and Wilson
coefficients can be considered as corrections which are, however, important for
achieving
better agreement with experimental data.

Hereafter we consider for simplicity only  the LO
approximation. 
The structure function
$F_2(x,Q^2)$ and PDFs have the following form (see \cite{Q2evo,HT})
\begin{eqnarray}
  F_2(x,Q^2) = e \, f_q(x,Q^2),
\label{8a}
\end{eqnarray}
where
$e= \sum\limits_{i = 1}^f e_i^2/f$ is an
average charge squared
with $f$ being a number of
active (massless) quark flavors.
The small-$x$ asymptotic expressions for sea quark and gluon densities $f_a(x, \mu^2)$
can be written as follows (both the LO and NLO results and their derivation can be found
\cite{Q2evo,Cvetic1}):
\begin{eqnarray}
f_a(x,\mu^2) &=&
f_a^{+}(x,\mu^2) + f_a^{-}(x,\mu^2), \nonumber \\
	f^{+}_g(x,\mu^2) &=& \biggl(A_g + C \,
A_q \biggl)
		\overline{I}_0(\sigma) \; e^{-\overline d_{+} s} + O(\rho),~~~C=\frac{C_F}{C_A}=\frac{4}{9}, \nonumber \\
f^{+}_q(x,\mu^2) &=&
\frac{\varphi}{3} \,\biggl(A_g + C \,
A_q \biggl)
\tilde{I}_1(\sigma)  \; e^{-\overline d_{+} s}
+ O(\rho), ~~~\varphi=\frac{f}{C_A}=\frac{f}{3},
\nonumber \\
        f^{-}_g(x,\mu^2) &=& - C \,
A_q e^{- d_{-} s} \, + \, O(x),~~
	f^{-}_q(x,\mu^2) ~=~  A_q e^{-d_{-} s} \, + \, O(x),
	\label{8.02}
\end{eqnarray}

\noindent
where $C_A=N_c$, $C_F=(N_c^2-1)/(2N_c)$ for the color $SU(N_c)$ group,
$\overline{I}_{\nu}(\sigma)$ and
$\tilde{I}_{\nu}(\sigma)$ ($\nu=0,1$)
are the combinations of the modified Bessel functions (at $s\geq 0$, i.e. $\mu^2 \geq Q^2_0$) and the usual
Bessel functions (at $s< 0$, i.e. $\mu^2 < Q^2_0$):
\begin{equation}
\tilde{I}_{\nu}(\sigma) =
\left\{
\begin{array}{ll}
\rho^{\nu} I_{\nu}(\sigma) , & \mbox{ if } s \geq 0; \\
(-\tilde{\rho})^{\nu} J_{\nu}(\tilde{\sigma}) , & \mbox{ if } s < 0.
\end{array}
\right. \, ,~~
\overline{I}_{\nu}(\sigma) =
\left\{
\begin{array}{ll}
\rho^{-\nu} I_{\nu}(\sigma) , & \mbox{ if } s \geq 0; \\
\tilde{\rho}^{-\nu} J_{\nu}(\tilde{\sigma}) , & \mbox{ if } s < 0.
\end{array}
\right.
\label{4}
\end{equation}
where $\overline{I}_{0}(\sigma) = \tilde{I}_{0}(\sigma)$ and
\bea
&&s=\ln \left( \frac{a_s(Q^2_0)}{a_s(\mu^2)} \right),~~
a_s(\mu^2) \equiv \frac{\alpha_s(\mu^2)}{4\pi} = \frac{1}{\beta_0\ln(\mu^2/\Lambda^2_{\rm LO})},~~
\sigma = 2\sqrt{\left|\hat{d}_+\right| s
  \ln \left( \frac{1}{x} \right)},~~ \nonumber \\
&&\rho=\frac{\sigma}{2\ln(1/x)},~~
\tilde{\sigma} = 2\sqrt{-\left|\hat{d}_+\right| s
  \ln \left( \frac{1}{x} \right)},~~ \tilde{\rho}=\frac{\tilde{\sigma}}{2\ln(1/x)}
\label{intro:1a}
\eea
and
\begin{equation}
\hat{d}_+ = - \frac{4C_A}{\beta_0} = - \frac{12}{\beta_0},~~~
\overline d_{+} = 1 + \frac{4f(1-C)}{3\beta_0} =
1 + \frac{20f}{27\beta_0},~~~
d_{-} = \frac{4Cf}{3\beta_0}= \frac{16f}{27\beta_0}
\label{intro:1b}
\end{equation}
are the singular and regular parts of the anomalous dimensions
and $\beta_0 = 11 -(2/3) f$ is the first coefficient of the QCD
$\beta$-function in the $\MSbar$-scheme.


The results for the parameters $A_a$ and $Q_0^2$ can be found in \cite{Cvetic1,Kotikov:2016oqm};
they were obtained\footnote{In the future, by using  (\ref{8a}) and (\ref{8.02}) and
  results of \cite{Illarionov:2008be}
we plan to perform the combined fits to the H1 and ZEUS experimental data\cite{37} and\cite{Abramowicz:1900rp} for
the DIS structure function $F_2(x,Q^2)$ and its charm part $F_2^c(x,Q^2)$, respectively.}
for $\alpha_s(M_Z)=0.1168$.

It is convenient to show the following expressions:
\begin{equation}
\beta_0 \, \hat{d}_+ = - 4C_A,~~~
\beta_0 \, \overline{d}_{+} = \frac{C_A}{3}\Bigl(11-2\varphi (1-2C)\Bigr),~~~
\beta_0 \, d_{-} = \frac{4Cf}{3}= \frac{4C_A \varphi}{3}.
\label{intro:1ba}
\end{equation}

\subsection{Kimber-Martin-Ryskin approach} \indent

The expressions~(3) --- (6) can be used as an input for the KMR procedure\cite{21}, giving
us the possibility to calculate the TMD parton density functions
in a proton.
Let $k_{\perp} \equiv k$, then the TMD parton distributions
in the differential $f_a^{(d)}(x,k^2,\mu^2)$ and integral $f_a^{(i)}(x,k^2,\mu^2)$ formulation
of the KMR approach can be written as
\bea
&&f^{(d)}_a(x,k^2,\mu^2) = \frac{\partial}{\partial \ln k^2} \Bigl[T_a(\mu^2,k^2) D_a(x,k^2)\Bigr] \, ,
\label{Def} \\
&&f^{(i)}_a(x,k^2,\mu^2) = T_a(\mu^2,k^2) \, \sum_{a'} \int\limits^{x_0}_x \, \frac{dz}{z} \,
P_{aa'}(z,k^2) \,  D_a\left(\frac{x}{z},k^2 \right) \, ,~~x_0=1-\Delta \, ,
\label{Def2}
\eea
where $D_a(x,\mu^2)$ are the conventional PDFs, $f_a(x,\mu^2) = x D_a(x,\mu^2)$, which obey
the DGLAP equations (see~(2.1) in\cite{Golec-Biernat:2018hqo}):
\be
\frac{\partial D_a(x,\mu^2)}{\partial \ln \mu^2} = \sum_{a'} \int\limits^{x_0}_x \, \frac{dz}{z} \,
P_{aa'}(z,\mu^2) \,  D_a\left(\frac{x}{z},\mu^2 \right) - D_a(x,\mu^2) \sum_{a'} \int\limits^{x_0}_0 \, dz \, z P_{a'a}(z,\mu^2)
\label{DGLAP}
\ee
with the splitting functions
\be
P_{aa'}(z,\mu^2) = 2a_s(\mu^2) \, P_{aa'}^{\rm (LO)} (z,\mu^2) + ...
\, .
\label{spliting}
\ee

\noindent
The Sudakov form factor $T_a(\mu^2,k^2)$ has the following form (see~(2.4) in\cite{Golec-Biernat:2018hqo}):
\be
T_a(\mu^2,k^2)= \exp \left\{ - \int\limits^{\mu^2}_{k^2} \, \frac{dp^2}{p^2} \,   \sum_{a'} \int\limits^{x_0}_0 \, dz \, z P_{a'a}(z,k^2) \right\}\, .
\label{Ta}
\ee

\section{Calculations} \indent

The DGLAP splitting functions at LO can be presented as (see, for example,~(2.56) --- (2.60) in\cite{Buras:1979yt})
\begin{equation}
 \displaystyle P_{qq}(z) = C_F \left[\frac{1+z^2}{(1-z)_+} + \frac{3}{2} \delta (1-z) \right], P_{qg}(z) = f \Bigl[z^2 + (1-z)^2\Bigr], \atop {
 \displaystyle P_{gq}(z) = C_F \left[\frac{1+(1-z)^2}{z}\right], \atop {
 \displaystyle P_{gg}(z) = 2C_A \left[\frac{z}{(1-z)_+} + \frac{1-z}{z} + z(1-z) + \frac{11-2\varphi}{12}  \delta (1-z) \right].}}
\label{splitingLO}
\end{equation}

\noindent
Since the upper limit of integrals in the r.h.s. of~(10) and (\ref{Ta}) is restricted by $x_0 \equiv 1-\Delta$,
the $\delta$-functions in~(13) give no contributions. Moreover, the "$+$" prescription is not needed: $(1-z)^{-1}_+ \to (1-z)^{-1}$.

\subsection{Sudakov form factors $T_a(\mu^2, k^2)$} \noindent

Evaluating~(\ref{Ta}), we have
\be
T_a(\mu^2,k^2) = \exp \Bigl[ -d_a R_a(\Delta) s_1 \Bigr] \, ,
\label{Ta.1}
\ee
where
\bea
&&s_1=\ln \left( \frac{a_s(k^2)}{a_s(\mu^2)} \right),~ d_a=\frac{4C_a}{\beta_0},~ C_q=C_F,~ C_g=C_A,~
\beta_0=\frac{C_A}{3} \Bigl(11-2\varphi\Bigr),~
\nonumber \\
&&R_q(\Delta)= \ln\left(\frac{1}{\Delta}\right) - \frac{3x_0^2}{4}
= \ln\left(\frac{1}{\Delta}\right) - \frac{3}{4} (1-\Delta)^2,~~ \nonumber \\
&&R_g(\Delta)= \ln\left(\frac{1}{\Delta}\right) - \left(1-\frac{\varphi}{4}\right) x_0^2
+ \frac{1-\varphi}{12} x_0^3 (4-3x_0) = \nonumber \\
&&= \ln\left(\frac{1}{\Delta}\right) - \left(1-\frac{\varphi}{4}\right) (1-\Delta)^2
+ \frac{1-\varphi}{12} (1-\Delta)^3 (1+3\Delta) \, .
\label{Ta.2}
\eea

\subsection{TMDs from differential formulation of KMR approach} \noindent

Now we can use (\ref{Def}) to find the results for TMD parton densities without derivatives.
Derivation of $T_a(\mu^2,k^2)$ is as follows
\be
\frac{\partial T_a(\mu^2,k^2)}{\partial \ln k^2} = d_a \, \beta_0 \, a_s(k^2) \, R_a(\Delta),
\label{DefTa}
\ee
\noindent
and derivations of conventional PDFs are as follows
\be
\frac{\partial f_a(x,k^2)}{\partial \ln k^2} = -\beta_0 \, a_s(k^2) \left[
\left(\frac{\hat{d}_+}{\rho_a} + \overline d_{+}\right) f_a^+(x,k^2) + d_{-} f_a^-(x,k^2) \right]
\, ,
\label{Deffa}
\ee
where
\be
\frac{1}{\rho_g} = \frac{\overline{I}_1(\sigma)}{\overline{I}_0(\sigma)},~~
 \frac{1}{\rho_q} = \frac{\overline{I}_0(\sigma)}{\tilde{I}_1(\sigma)}.
\label{rho.a}
\ee
\noindent
So, the result for the TMD parton densities reads
\begin{equation}
 \displaystyle f^{(d)}_a(x,k^2,\mu^2) = \beta_0 \, a_s(k^2) T_a(\mu^2,k^2) \times \atop {
 \displaystyle \times \left(d_a \, R_a(\Delta) \, f_a(x,k^2) - \left[\left(\frac{\hat{d}_+}{\rho_a} + \overline d_{+}\right) f_a^+(x,k^2) + d_{-} f_a^-(k^2) \right] \right) = 4C_a \, a_s(k^2) \times \atop {
 \displaystyle \times T_a(\mu^2,k^2) \left(R_a(\Delta) \, f_a(x,k^2) - \frac{1}{d_a} \left[\left(\frac{\hat{d}_+}{\rho_a} + \overline d_{+}\right) f_a^+(x,k^2) + d_{-} f_a^-(k^2) \right] \right). }}
\label{uPDF}
\end{equation}

\noindent
The expressions for conventional PDFs and Sudakov form factors are given in the Sections~2 and~3.1, respectively.
Since the value of $\hat{d}_+$ is negative and the factor $\hat{d}_+/\rho_a$ is large at low $x$, the TMDs $f^{(d)}_a(x,k^2,\mu^2)$ are
positive at small $x$. With increasing $x$, the results
for the latter can be negative.

\subsection{TMDs from integral formulation of KMR approach} \noindent

Now we can use~(\ref{Def2}) to find the results for TMDs without derivatives.
After some algebra we have
\begin{equation}
 \displaystyle f^{(i)}_a(x,k^2,\mu^2) = 4C_a \, a_s(k^2) T_a(\mu^2,k^2) \times \atop {
 \displaystyle \times \left(D_a(\Delta) \, f_a\left(\frac{x}{x_0},k^2\right) + D_a^+ f_a^+\left(\frac{x}{x_0},k^2\right) + D_a^{-} f_a^-\left({x\over x_0}, k^2\right) \right) = \atop {
 \displaystyle = 4C_a \, a_s(k^2) T_a(\mu^2,k^2) \left(\overline{D}_a(\Delta) \, f_a\left(\frac{x}{x_0},k^2\right) + \overline{D}_a^+ f_a^+\left(\frac{x}{x_0},k^2\right) \right), }}
\label{uPDF2.1}
\end{equation}

where
\be
\overline{D}_a(\Delta) = D_a(\Delta) + D_a^{-},~~\overline{D}_a^+ = D_a^+ -D_a^{-} \, .
\label{uPDF2.2}
\ee

Using the relations from~(\ref{intro:1ba}), we can obtain:
\bea
&&D_q(\Delta)= \ln\left(\frac{1}{\Delta}\right) - \frac{x_0}{4} (2+x_0)
= \ln\left(\frac{1}{\Delta}\right) - \frac{1-\Delta}{4} (3-\Delta), \nonumber \\
&&D_g(\Delta)= \ln\left(\frac{1}{\Delta}\right) - x_0 + \frac{x_0^2}{4} - \frac{x_0^3}{3}
= \ln\left(\frac{1}{\Delta}\right) -
 \frac{1-\Delta}{12} (13-5\Delta + 4\Delta^2),~~  \nonumber \\
&&D_q^-(\Delta) = - \frac{x_0\varphi}{2} \,  \left(1-x_0 + \frac{2x_0^2}{3}\right)
= - \frac{(1-\Delta)\varphi}{6} \,  (2-\Delta + 2\Delta^2),~~ D_g^-(\Delta)=0, \nonumber \\
&&D_g^+= \frac{1}{\overline{\rho}_g} - x_0 + \frac{x_0^2}{4}
+ \frac{C\varphi}{3}, \,
\nonumber \\
&&D_q^+=  \frac{3x_0}{2C} \Biggl[\frac{1}{\overline{\rho}_s} \left(1 - x_0  + \frac{2x_0^2}{3}\right) -
\left( 1- \frac{x_0}{2} + \frac{2x_0^2}{9}  \right)\Biggr],
\label{uPDF2.3}
\eea
where
\be
\overline{\sigma} =
\left\{
\begin{array}{ll}
\sigma\left(x \to \frac{x}{x_0}\right) , & \mbox{ if } s \geq 0; \\
\tilde{\sigma}\left(x \to \frac{x}{x_0}\right) , & \mbox{ if } s < 0.
\end{array}
\right. \, ,~~
\frac{1}{\overline{\rho}_a} = \frac{1}{\rho}_a \left(x \to \frac{x}{x_0}\right) \, .
\label{rho}
\ee

\subsection{Cut-off parameter $\Delta$} \noindent

For the phenomenological applications, the cut-off parameter $\Delta$ usually has one of two basic forms:
\be
\Delta_1=\frac{k}{\mu},~~\Delta_2=\frac{k}{k+\mu},
\label{Delta12}
\ee

\noindent
that reflects the two cases: $\Delta_1$ is in the strong ordering, $\Delta_2$ is in the angular ordering (see \cite{Golec-Biernat:2018hqo}).
In all above cases, except the results for $T_a(\mu^2,k^2)$, we can simply replace the parameter $\Delta$
by
$\Delta_1$ and/or $\Delta_2$. For the Sudakov form factors, we note
that the parameters $\Delta_i$ (with $i=1, 2$) contribute to the integrand
in (\ref{Ta}) and, thus, their
momentum dependence changes the results in (\ref{Ta.1}). To perform the correct evaluation of the integral (\ref{Ta}),
we should
recalculate the $p^2$ integration in (\ref{Ta}). So, we have
\be
T_a^{(i)}(\mu^2,k^2) = \exp \left[ -4C_a \int\limits^{\mu^2}_{k^2} \, \frac{dp^2}{p^2} \, a_s(p^2) R_a(\Delta_i)  \right].
\label{Ta.2}
\ee

\noindent
The analytic evaluation of $T_a^{(i)}(\mu^2,k^2)$ is a very cumbersome procedure,
which will be
accomplished in the future. With the purpose of simplifying
our analysis, below we use the numerical
results for $T_a^{(i)}(\mu^2,k^2)$.

\subsection{Comparison of $f^{(i)}_a(x,k^2,\mu^2)$ and $f^{(d)}_a(x,k^2,\mu^2)$} \noindent

To perform the comparison between the two obtained expressions,
it is convenient to rewrite~(\ref{uPDF}) as
\begin{equation}
  \displaystyle f^{(1)}_a(x,k^2,\mu^2) = 4C_a \, a_s(k^2) T_a(\mu^2,k^2) \times \atop {
  \displaystyle \times \left(R_a(\Delta) \, f_a(x,k^2) + t_a^+ f_a^+(x,k^2) + t_a^{-} f_a^-(x,k^2) \right) = \atop {
  \displaystyle = 4C_a \, a_s(k^2) T_a(\mu^2,k^2) \left(\overline{R}_a(\Delta) \, f_a(x,k^2) + \overline{t}_a^+ f_a^+(x,k^2) \right)}},
\label{uPDF1.1}
\end{equation}

\noindent
where
\be
t_a^+ = -\frac{1}{d_a} \, \left(\frac{\hat{d}_+}{\rho_a} + \overline d_{+}\right),~~
t_a^- = -\frac{d_{-}}{d_a},~~
~~\overline{R}_a(\Delta) = R_a(\Delta) + t_a^{-},~~\overline{t}_a^+ = t_a^+ -t_a^{-}.
\label{uPDF.2}
\ee

\noindent
Using the relations from~(\ref{intro:1ba}), we can obtain:
\begin{equation}
  \displaystyle t_g^+= \frac{1}{\rho_g} - \frac{1}{12} \Bigl(11+2\varphi (1-2C)\Bigr),~~ t_g^-= -\frac{C\varphi}{3}, \atop {
  \displaystyle \overline{t}_g^+= \frac{1}{\rho_g} - \frac{1}{12} \Bigl(11+2\varphi (1-4C)\Bigr),~~t_q^{\pm}= \frac{1}{C} t_g^{\pm}(\rho_g \to \rho_q).}
\label{uPDF.3}
\end{equation}

\noindent
It is convenient to compare the results for $f^{(i)}_a(x,k^2,\mu^2)$ and $f^{(d)}_a(x,k^2,\mu^2)$, i.e. the values
 of $\overline{D}_a(\Delta)$, $\overline{D}_a^+(\Delta)$ and  $\overline{R}_a(\Delta)$, $\overline{t}_a^+(\Delta)$
 at $\Delta \to 0$, when $k^2 \ll \mu^2$:
\bea
&&\overline{R}_q(\Delta) = \ln\left(\frac{1}{\Delta}\right) - \frac{3}{4} - \frac{\varphi}{3},~~
\overline{R}_g(\Delta) = \ln\left(\frac{1}{\Delta}\right) - \frac{11}{12} + \frac{\varphi(1-2C)}{6}, \nonumber \\
&&\overline{D}_q(\Delta) = \ln\left(\frac{1}{\Delta}\right) - \frac{3}{4} - \frac{\varphi}{3},~~
\overline{D}_g(\Delta) = \ln\left(\frac{1}{\Delta}\right) - \frac{13}{12}, \nonumber \\
&&\overline{t}_q^+= \frac{1}{C} \left(\frac{1}{\rho_q} - \frac{11}{12} - \frac{\varphi(1-4C)}{6}\right),~~
\overline{t}_g^+= \frac{1}{\rho_g} - \frac{11}{12} - \frac{\varphi(1-4C)}{6}, \nonumber \\
&&\overline{D}_q^+= \frac{1}{C} \left(\frac{1}{\rho_q} - \frac{13}{12} + \frac{C\varphi}{3}\right),~~
\overline{t}_g^+= \frac{1}{\rho_g} - \frac{3}{4} + \frac{C \varphi}{3},
\label{uPDF2.4}
\eea
or
\bea
&&\overline{R}_q(\Delta) =\overline{D}_q(\Delta),~~ \overline{R}_g(\Delta) =\overline{D}_g(\Delta) + \frac{1}{6} \Bigl(
1+\varphi(1-2C) \Bigr), \nonumber \\
&&\overline{t}_q^+ =\overline{D}_q^+ + \frac{1}{6} \Bigl(1-\varphi(1-2C) \Bigr),~~ \overline{t}_g^+ =\overline{D}_g^+ - \frac{1}{6} \Bigl( 1+\varphi(1-2C) \Bigr).
\label{uPDF2.5}
\eea

\noindent
So, the difference is regular at $\Delta \to 0$ and it does not give large contributions.
This is clearly illustrated in Fig.~1, where we show the ratio of gluon densities
$f^{(i)}_a(x,k^2,\mu^2)$ and $f^{(d)}_a(x,k^2,\mu^2)$ calculated with
angular ordering condition applied
as a function of $k_T^2 \equiv k^2$ at several values of $x$ and $\mu^2$.
As one can see, with increasing  $k^2$ the difference between
these two approaches becomes more pronounced.

\subsection{Infrared modification of the strong coupling} \noindent

The equations~(\ref{8.02}) at $s<0$ were
used in\cite{HT}, where
the higher-twist corrections through twist six
were added to find good agreement with the experimental data for the deep inelatic proton structure
function $F_2(x,Q^2)$ for $Q^2 \geq 0.5$~GeV$^2$.
However, such application is not so useful here
because the case with  $s<0$
may lead to the negative TMDs and, hence, to the negative
cross sections of the physical processes.

To overcome these problems, which emerge
at small $k^2$ values,
we investigate an alternative possibility following to\cite{Cvetic1}; namely, the modification of the
strong-coupling constant in the infrared region.
Specifically, we consider two modifications, which effectively increase the
argument of the strong coupling constant at small $\mu^2$ values, in accordance
with\cite{Kotikov:1993yw,Andersson:2002cf}.
In the first case, which is more phenomenological, we introduce a freezing
of the strong-coupling constant by changing its argument as
$\mu^2 \to \mu^2 + M^2_{\rho}$, where $M_{\rho}$ is the $\rho$ meson mass
\cite{Badelek:1996ap}.
Thus, in the formulae of Section~3
we introduce the following replacement
\begin{equation}
\alpha_s(\mu^2) \to \alpha_{\rm fr}(\mu^2)
= \alpha_s(\mu^2 + M^2_{\rho}) \, .
\label{Intro:2}
\end{equation}

\noindent
The second possibility is based on the idea by Shirkov and Solovtsov
\cite{Shirkov:1997wi} (see also the recent reviews\cite{Cvetic:2008bn} and the references
therein) regarding the analyticity of the strong coupling that
leads to an additional power dependence.
In this case,
the
QCD coupling
$\alpha_s(\mu^2)$
appearing
in the formulae of the previous Sections is to
be replaced as
\be
	\alpha_s(\mu^2) \to \alpha_{\rm an}(\mu^2) =
		\alpha_s(\mu^2) - \dfrac{1}{\beta_0}\,
\dfrac{\Lambda^2_{\rm LO}}{\mu^2 - \Lambda^2_{\rm LO}} \, .
\label{an:NLO}
\ee

\noindent
Such replacements have been done\cite{Cvetic1}, where we took the normalizations magnitudes $A_g$ and $A_q$.
As we can see from\cite{Cvetic1,Kotikov:2016oqm},
the fits based on the frozen and
analytic strong-coupling constants are very similar and describe the $F_2(x,Q^2)$ data in
the small-$Q^2$ range significantly better than the canonical fit.

\subsection{Beyond small $x$} \noindent

In the phenomenological applications (see Section~4) the calculated TMD parton
densities will be used to predict the cross sections of several high-energy processes.
According to $k_T$-factorization approach\cite{3,4}, the theoretical predictions for the cross sections
can be obtained by convolution of these TMD parton densities
and the corresponding off-shell production amplitudes. So, we need the TMD
quark and gluon distributions
in rather broad range of the $x$ variable, i.e. beyond the standart low $x$ range ($x \leq 0.05$).

Our TMD parton densities are
exactly expressed through the conventional PDFs $f_a(x,\mu^2)$ as it was shown in the equations (\ref{uPDF1.1}) and (\ref{uPDF2.1}).
Then,
the densities  $f_a(x,\mu^2)$ listed in Section 2.1 should be extended
in the following form\cite{Gross:1974fm,Lopez:1979bb}
(see, for example, the recent paper \cite{Kotikov:2018cju}, where similar extension has been done in the case of
EMC effect from the study of shadowing \cite{Kotikov:2017mhk} at low $x$ to antishadowing effect at $x \sim 0.1 - 0.2$):
\be
{f}_a(x,\mu^2) \to f_a(x,\mu^2) \, (1-x)^{\beta_a(s)}, ~~ \beta_a(s) = \beta_a(0) + \frac{4C_a s}{\beta_0}.~~
\label{Large}
 \ee
\noindent
Note that such form was successfully used in the conventional PDF parametrizations (see\cite{Lopez:1979bb,Illarionov:2010gy}).
The value of $\beta_a(0)$ can be estimated from the quark counting rules\cite{Matveev:1973ra}:
\be
\beta_v(0) \sim 3,~~  \beta_g(0) \sim \beta_v(0)+1 \sim 4,~~  \beta_q(0) \sim \beta_v(0)+2 \sim 5 \, ,
\label{bV}
 \ee
where the symbol $v$ marks the valence part of quark density.
Usually the $\beta_v(0)$, $\beta_g(0)$, $\beta_q(0)$ are determined from fits of experimental data
(see, for example,\cite{Jimenez-Delgado:2014twa,Kotikov:2016ljf,Krivokhizhin:2005pt}) and the results for these values may be quite different,
because various groups producing PDF sets use
different sets of experimental data or take some privilege for their parts.
Moreover,
the difference can be attributed to various choices
of the initial condition $\mu^2_0$ of the  $\mu^2$-evolution
but it
should be not so strong because the $\mu^2$-dependence is double-logarithmic.

It is
convenient to assume that similar relations take place just beyond the standard
low $x$ range ($x \leq 0.05$).
Thus, the TMD parton densities can be modified in the form, similar to~(33),
that leads to
\bea
f^{(d)}_a(x,k^2,\mu^2) \to f^{(d)}_a(x,k^2,\mu^2) \, (1-x)^{\beta_a(s)}, \label{uPDF1.over1} \\
f^{(i)}_a(x,k^2,\mu^2) \to f^{(i)}_a(x,k^2,\mu^2) \, \left(1-\frac{x}{x_0}\right)^{\beta_a(s)}. \label{uPDF2.over1}
\eea

\noindent
In our analysis, the numerical values of $\beta_{g}(0)$ have been extracted from the fit
to the inclusive $b$-jet production data taken by the CMS\cite{59} and ATLAS\cite{60} Collaborations
in $pp$ collisions at $\sqrt s = 7$~TeV (see Section~4.1 below). We find
that best description of the leading $b$-jet transverse momentum distributions
in a whole kinematical region is achieved with $\beta_{g}(0) = 3.03$ and $\beta_{g}(0) = 5.77$ for
"frozen" and analytic strong coupling constant~(37) and (38), respectively.
We see that the obtained results are close to ones in (\ref{bV}).

The TMD gluon densities in a proton
obtained with appropriate treatment of the
strong coupling and $\beta_g(0)$ are shown in Fig.~1 as a function of
transverse momentum $k_T^2$ for different values of proton
longitudinal momentum fraction $x$ and hard scale $\mu^2$.
We have used the integral formulation of KMR procedure as given by~(36)
for illustration. The solid green and yellow curves correspond to the
results obtained with "frozen" and analytic coupling constant with
angular ordering condition, while corresponding dashed curves
represent the results obtained with strong ordering condition.
As one can see, the strong ordering condition leads to
a steep drop of the gluon densities beyond the scale $\mu^2$.
It contrasts with angular ordering, where the gluon transverse momentum is
allowed to be larger than $\mu^2$ (see\cite{Golec-Biernat:2018hqo}).
We also show here the TMD gluon distributions
calculated numerically in the traditional KMR scenario, where the
conventional parton densities from standard MMHT'2014 (LO) set\cite{61} were used as an input (red curves on Fig.~1).
The results of our analytical calculations are nicely agree with
the latter for $k_T^2 \geq 10$~GeV in wide $x$ region (up to $x \leq 0.05$), that demonstrates the
applicability of the generalized DAS approximation.
At $k_T^2 < \mu_0^2 \sim 1$~GeV$^2$
the numerically calculated KMR gluon density is modelled to be a flat
according to the prescription\cite{21} under strong normalization condition
\bea
  \int\limits_0^{\mu^2} f_a(x, k^2, \mu^2) dk^2 = f_a(x, \mu^2),
\eea

\noindent
which is often used in the KMR scheme.
Such determination, of course, leads to a low $k_T^2$ plateau, clearly seen in Fig.~1. In contrast, our formalism
with appropriate modifications of strong coupling as described above
results in continuous TMD quark and gluon density functions, well defined in a whole $k_T^2$ region.
Below we will consider the
phenomenological consequences of our approach.

\section{Phenomenological applications} \noindent

We are now in a position to apply the obtained TMD parton densities in a proton to
several hard QCD processes studied at hadron colliders. In the present paper we
consider the inclusive production of $b$-jets and Higgs bosons at the LHC conditions
and charm and beauty contributions to the deep inelastic proton structure function $F_2(x,Q^2)$
measured in $ep$ collisions at HERA.
These processes have been already investigated within the
$k_T$-factorization approach
and found to be strongly sensitive to the gluon
content of the proton.
To calculate the total and differential cross sections of $b$-jets, Higgs boson production
and proton structure functions $F_2^c(x,Q^2)$ and $F_2^b(x, Q^2)$ we strictly follow our
previous considerations\cite{11,12,15,16,62}. Everywhere below we have used
one-loop formula for the strong coupling constant with
$n_f = 4$ active quark flavors and $\Lambda_{\rm QCD} = 143$~MeV (that corresponds to
$\alpha_s(m_Z^2) = 0.1168$) for analytically calculated TMD quark and gluon
densities as described above and apply $n_f = 5$ with $\alpha_s(m_Z^2) = 0.13$
for KMR partons evaluated numerically. The latter choice is dictated by the parameter setup
employed in the MMHT'2014 (LO) PDFs\cite{61}, used here as an input for KMR procedure.

\subsection{Inclusive $b$-jet production at the LHC} \noindent

Following\cite{15,16}, our consideration is based on the leading
off-shell (depending on the transverse momenta of
incoming particles) gluon fusion subprocess $g^* (k_1) + g^*(k_2) \to b (p_1) + \bar b(p_2)$,
where the four-momenta of all particles are indicated in parentheses.
According to the $k_T$-factorization prescription\cite{3,4}, the corresponding cross section
can be written as
\begin{equation}
  \displaystyle \sigma = \int dx_1 dx_2 \int d{\mathbf k}_{1T}^2 d{\mathbf k}_{2T}^2 f_g(x_1,{\mathbf k}_{1T}^2,\mu^2) f_g(x_2,{\mathbf k}_{2T}^2,\mu^2) \times \atop {
  \displaystyle \times d\sigma^*(x_1,x_2,{\mathbf k}_{1T}^2,{\mathbf k}_{2T}^2,\mu^2)},
\end{equation}

\noindent
where $\sigma^*(x_1,x_2,{\mathbf k}_{1T}^2,{\mathbf k}_{2T}^2,\mu^2)$ is the
off-shell partonic cross section and ${\mathbf k}_{1T}^2$ and ${\mathbf k}_{2T}^2$ are the non-zero
two-dimensional transverse momenta of incoming partons. The detailed description of the calculation steps
(including the evaluation of the off-shell amplitudes) can be found in\cite{15,16}.
Here we only specify the essential numerical parameters. So, following\cite{63},
we set the $b$-quark mass $m_b = 4.78$~GeV and, as it often done in pQCD calculations,
choose the default renormalization and factorization scales $\mu_R$ and $\mu_F$ to be equal
to leading $b$-jet transverse momentum.
The calculations were performed using newly developed Monte-Carlo event generator \textsc{pegasus}\cite{64}.

The CMS Collaboration has measured the double differential cross section $d\sigma/dp_T dy$
of inclusive $b$-jet production at  $\sqrt s = 7$~TeV
in five $b$-jet rapidity regions, namely,
$|y| < 0.5$, $0.5 < |y| < 1$, $1 < |y | < 1.5$, $1.5 < |y| < 2$ and $2 < |y| < 2.2$ as a function of
the leading $b$-jet transverse momentum\cite{59}.
In the ATLAS analysis\cite{60}, the inclusive $b$-jet cross section
has been measured as a function of transverse momentum $p_T$ in the range
$20 < p_T < 400$~GeV and rapidity in the range $|y| < 2.1$. In addition, the $b\bar b$-dijet
cross section has been measured as a function of the dijet invariant mass $M$ in the
range $110 < M < 760$~GeV, azimuthal angle difference $\Delta \phi$ between the two $b$-jets
and angular variable $\chi = \exp |y_1 - y_2|$ for jets with $p_T > 40$~GeV
in two dijet mass regions.

The results of our calculations are shown in Figs.~3 --- 5 in comparison with the CMS and ATLAS data\cite{59,60}.
The solid green and yellow histograms were obtained with the TMD gluon density as given by~(20) --- (23)
with "frozen" and analytic QCD coupling by fixing both the renormalization and factorization scales at their
default values.
The red histograms represent the results obtained with
the numerically calculated KMR gluon distributions.
The shaded bands correspond to scale uncertainties of these predictions.
As usual, the latter
have been estimated by varying the scales $\mu_R$ and $\mu_F$ by a factor
of $2$ around their default values.
We have obtained a good description of the $b$-jet transverse momentum distributions
in each of the rapidity subdivisions, both in normalization and the shape.
Our predictions are only tend to slightly underestimate the measured cross sections
at very high transverse momenta $p_T \sim 200 - 400$~GeV,
but they agree with the data within
the theoretical and experimental uncertainties.
The results obtained with numerically calculated KMR gluon density
agree with data and analytical TMD gluons at low and moderate transverse momenta, but
overestimate both the CMS and ATLAS data at $p_T > 100$~GeV,
especially at forward rapidities.
We note that here the essentially large $x$
region is probed, so the better description of the data
achieved with analytical TMD gluon distributions demonstrates
that their large-$x$ extension, as described above in Section~3.7, is rather reasonable.

All the considered TMD gluons show good agreement with
the $b\bar b$-dijet cross sections measured by the ATLAS Collaboration.
In particular, the good description of the $\Delta \phi$ distribution
is remarkable, since the latter is known to be a strongly sensitive
to the $k_T^2$ shape of the TMD gluon density (see \cite{15,16} and references therein).
As it was expected, the $\chi$ distribution flattens for large
invariant masses $M$. Note that here an additional acceptance
requirement, that restricts the boost of the dijet system to $|y_{\rm boost}| = |y_1 + y_2|/2 < 1.1$,
has been applied for $\chi$ measurements. This requirement significantly reduces\cite{60} the sensitivity to
gluon density function at small $x$ and all
theoretical predictions for $\chi$ distributions are practically coincide.
Thus, we conclude that our analytical TMD
parton densities given by~(20) --- (23) does not
contradict available LHC data on $b$-jet production.

\subsection{Inclusive Higgs boson production at the LHC} \noindent

Our consideration is mainly based on the off-shell amplitude of the gluon-gluon fusion
subprocess $g^*(k_1) + g^*(k_2) \to H(p)$ calculated using the effective Lagrangian\cite{65,66}
for the Higgs coupling to gluons and extended recently to the subsequent
$H \to \gamma \gamma$, $H \to ZZ^* \to 4l$ (where $l = e$ or $\mu$)
and $H \to W^+W^- \to e^\pm \mu^\mp \nu \bar \nu$ decays.
The details of the calculations
are explained in\cite{11,12} and here we strictly follow our previous consideration.
Everywhere below, we set the Higgs boson mass $m_H = 125.1$~GeV and
its full decay width $\Gamma_H = 4.3$~MeV.
The default values of the renormalization and factorization scales are chosen to be equal to
Higgs mass. The cross sections were produced with Monte-Carlo generator \textsc{pegasus}\cite{64}.

The latest measurements of the inclusive Higgs boson production (in the diphoton decay mode)
were performed by the CMS\cite{67} and ATLAS\cite{68} Collaborations at the LHC energy $\sqrt s = 13$~TeV.
In the CMS analysis, two isolated final state photons originating from the Higgs boson decays are required to have
pseudorapidities $|\eta^\gamma| < 2.5$, excluding the region $1.4442 < |\eta^\gamma| < 1.566$.
Additionally, photons with largest and
next-to-largest transverse momentum $p_T^\gamma$ (so-called leading and subleading photons)
must satisfy the conditions
of $p_T^\gamma/M^{\gamma \gamma} > 1/3$ and $p_T^\gamma/M^{\gamma \gamma} > 1/4$ respectively,
where $M^{\gamma \gamma}$ is the diphoton pair mass, $M^{\gamma \gamma} > 90$~GeV.
In the ATLAS measurement\cite{68} both of these decay photons must have pseudorapidities
$|\eta^\gamma| < 2.37$ (excluding $1.37 < |\eta^\gamma| < 1.52$) with the leading (subleading) photon satisfying
$p_T^\gamma/M^{\gamma \gamma} > 0.35$ and $p_T^\gamma/M^{\gamma \gamma} > 0.25$,
while invariant mass $M^{\gamma \gamma}$ is required to be $105 < M^{\gamma \gamma} < 160$~GeV.
We have implemented experimental setup in our numerical program.
The Higgs transverse momentum $p_T$, absolute value of the rapidity $y$ and cosine of
photon helicity angle $\cos \theta^*$ (in  the  Collins-Soper  frame) were measured\cite{67,68}.
Both $p_T$ and $y$ probe the production mechanism and parton distribution functions in a proton,
while $\cos \theta^*$ is related to spin-CP nature of the decaying Higgs boson.

The results of our calculations are shown in Figs.~6 and 7 in comparison with latest LHC
data. One can see that our predictions with both analytic and "frozen" treatment of QCD coupling
reasonably agree with the data for all considered kinematical observables, although some tendency to slightly
overestimate the LHC data in the low $p_T$ region is observed.
This tendency results in a some ovestimation of the rapidity and
photon helicity angle distributions, but the predictions are still agree
with the data within the experimental and theoretical uncertainties,
calculated as it was described above.
The scale dependence of our predictions, of course, exceeds the uncertainties of conventional
higher-order pQCD calculations (which are about of 10 --- 11\%).
However, it could be easily
understood because only the tree-level LO hard scaterring amplitudes are involved.
The strong drop in the $|\cos \theta^*|$ distribution around $|\cos \theta^*| \sim 0.6$
is due to the fiducial requirement on the photon system
originating from the scalar Higgs boson decay.
The calculations based on the numerically evaluated KMR gluon density
agree well with the data at low $p_T$ and tend to overshoot them at high
transverse momenta.
Note that we added to our results contributions
from weak boson fusion ($W^+W^- \to H$ and $ZZ \to H$),
associated $HZ$ or $HW^\pm$ production and associated $t\bar t H$
production (grey shaded bands in Figs.~6 and 7). These contributions
are essential at high $p_T$ and have been calculated in the conventional
pQCD approach with the NLO accuracy. We take them from\cite{67,68}.
Once again, we can conclude that the analytical expressions for TMD parton densities~(20) --- (23)
does not contradict available LHC data in the probed kinematical region, where $\mu^2 \sim m_H^2$.

\subsection{Proton structure functions $F_2^c(x,Q^2)$ and $F_2^b(x,Q^2)$} \noindent

The important information on the quark and gluon structure of proton can be also extracted
from the data on deep inelastic $ep$ scattering. Its differential cross section can be presented in
the simple form:
\begin{equation}
  {d^2 \sigma \over dx dy} = {2 \pi \alpha^2 \over x Q^4} \left[ \left( 1 - y + {y^2\over 2}\right) F_2(x,Q^2) - {y^2 \over 2} F_L(x, Q^2)\right],
\end{equation}

\noindent
where $F_2(x, Q^2)$ and $F_L(x, Q^2)$ are the proton transverse and longitudinal structure functions,
$x$ and $y$ are the usual Bjorken scaling variables.
The charm and beauty contributions to $F_2(x,Q^2)$ are described through
perturbative production of charm or beauty quarks and, therefore, directly related with the
gluon content of the proton. Our evaluation below is based on the formulas\cite{62}
and here we again strictly follow our previous consideration in all aspects.
We only note that the charm and beauty masses are set to be equal to
$m_c = 1.65$~GeV and $m_b = 4.78$~GeV\cite{63}.

Our results are shown in Figs.~8 and~9 in comparison with the latest ZEUS\cite{69} and
H1\cite{70,71} data.
We find that the predictions obtained with "frozen" strong coupling are
in perfect agreement with the latest HERA data
for both structure functions $F_2^c(x,Q^2)$
and $F_2^b(x,Q^2)$ in a wide region of $x$ and $Q^2$,
both in normalization and shape. These predictions slightly overshoot the ones
obtained with analytic treatment of the QCD coupling constant. The difference between
these two approaches becomes more clearly pronounced at small $x$ and low $Q^2$ values,
$Q^2 \leq 10$~GeV$^2$.
The predictions based on the analytic QCD coupling
still agree with the data on $F_2^b(x,Q^2)$ within the theoretical and
experimental uncertainties, but clearly underestimate the data on $F_2^c(x,Q^2)$,
especially at low $Q^2$. However, we note that some reasonable variation
in charmed quark mass $m_c = 1.65 \pm 0.2$~GeV can almost
eliminate the visible disagreement (not shown in Figs.~8 and 9).
The traditional KMR approach underestimates the HERA data on both $F_2^c(x,Q^2)$ and $F_2^b(x,Q^2)$
at small $x$ and relatively low $Q^2 \leq 30 - 60$~GeV$^2$, where
the higher-order QCD corrections are known to be important. The difference between
all the theoretical predictions becomes negligible
with increasing of $Q^2$.
Thus, we can conclude that the proposed analytical calculations of the
TMD parton densities in a proton does not contradict the latest HERA data,
although the best description of the latter (with the default parameter set)
is achieved with "frozen" QCD coupling constant.

\section{Conclusions} \noindent

We presented the analytical calculations of the
transverse momentum dependent parton densities in a proton. These calculations
are based on the Bessel-inspired behavior of parton densities at small Bjorken $x$,
obtained in the case of the flat initial conditions for DGLAP evolution equations
in the double scaling QCD approximation.
To construct the TMD parton distributions we applied the leading-order Kimber-Martin-Ryskin approach,
which is widely used in the phenomenological applications. We implemented the different
treatments of kinematical constraint, reflecting the angular and strong ordering conditions
and discussed the relations between the differential and integral formulation of the KMR approach.
Finally, we demonstrated that the calculated TMD parton distributions does
not contradict the LHC data on inclusive $b$-jet production at $\sqrt s = 7$~TeV,
inclusive Higg boson production (in diphoton decay mode) at $\sqrt s = 13$~TeV and
latest HERA data on the charm and beauty contributions to the
deep inelastic proton structure function $F_2(x,Q^2)$ in a wide region of $x$ and $Q^2$.

As the next step, we plan to study the longitudinal
structure function $F_L(x,Q^2)$ and also to provide predictions for its heavy quark parts, $F_L^c(x,Q^2)$ and $F_L^b(x,Q^2)$,
which will be compare with  our other predictions\cite{62}.
Moreover, we plan to extend the present analysis beyond the LO approximation.
We will obtain the results for the NLO TMD parton densities using the corresponding NLO results\cite{Cvetic1,Q2evo,HT}
for the standard PDFs in the generalized DAS approach. We will accept also the results for the NLO matrix elements
(see\cite{72,25} and references and discussions therein).

\section*{Acknowledgements}
We thank H.~Jung, S.P.~Baranov and M.A.~Malyshev for very useful discussions
and remarks.
A.V.K. highly appreciates the warm hospitality at the Institute of
Modern Physics CAS (Lanzhou, China) and thanks the CAS President’s International Fellowship Initiative (Grant No. 2017VMA0040) for support.
A.V.L. is grateful to Institute of Modern Physics CAS (Lanzhou, China) for support
and warm hospitality and DESY Directorate for the support in the framework of
Cooperation Agreement between MSU and DESY on phenomenology of the LHC processes
and TMD parton densities.
 P.Z. is supported in part by the National Natural Science Foundation of China
(Grants No. 11975320).

\newpage

\begin{figure}
\begin{center}
\includegraphics[width=7.9cm]{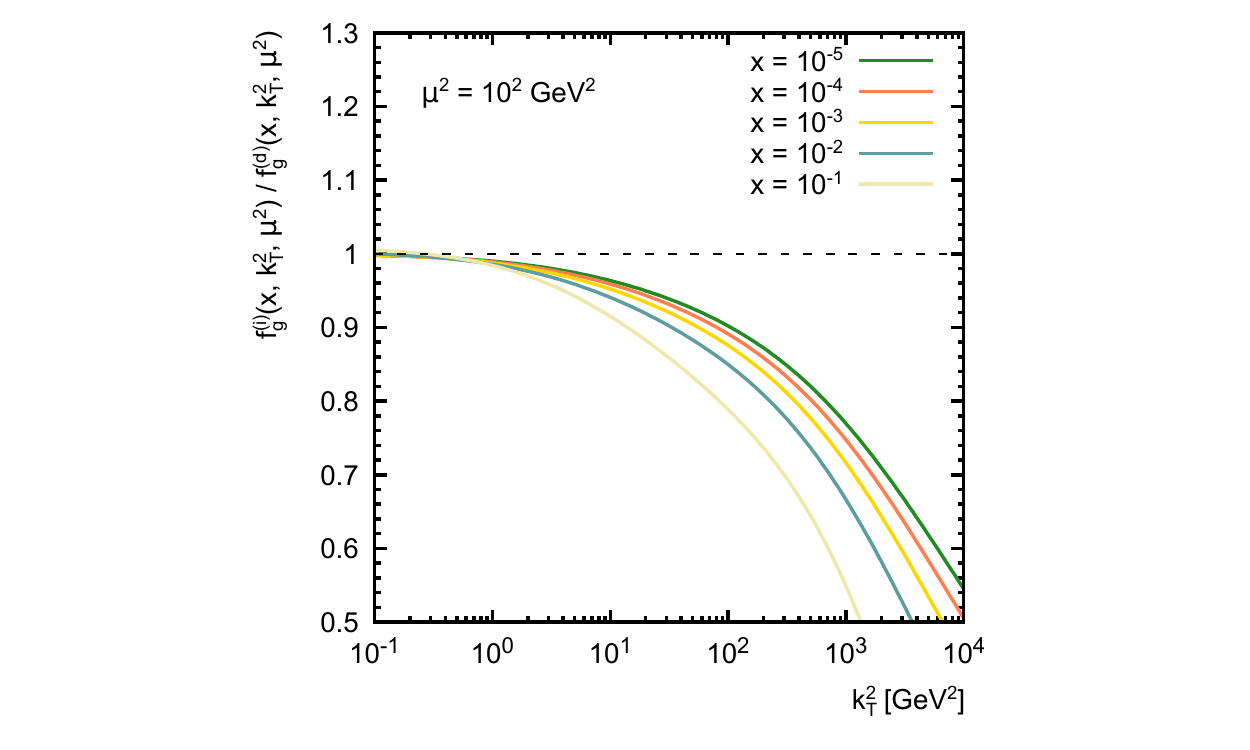}
\includegraphics[width=7.9cm]{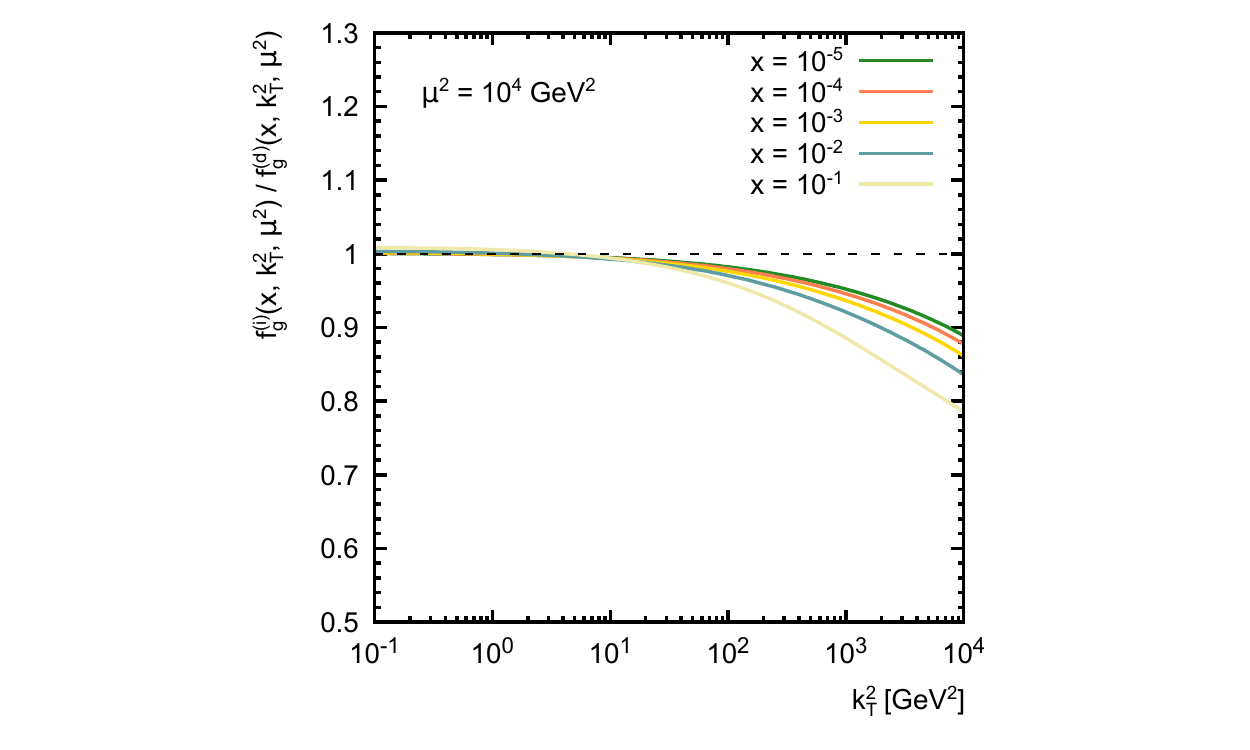}
\caption{The ratio of the TMD gluon densities in a proton
obtained using integral and differential formulation of the KMR approach
as a function of the gluon transverse momentum $k_T^2$ at different values of
longitudinal momentum fraction $x$ and hard scale $\mu^2$.
The angular ordering condition was applied.}
\label{fig1}
\end{center}
\end{figure}

\begin{figure}
\begin{center}
\includegraphics[width=7.9cm]{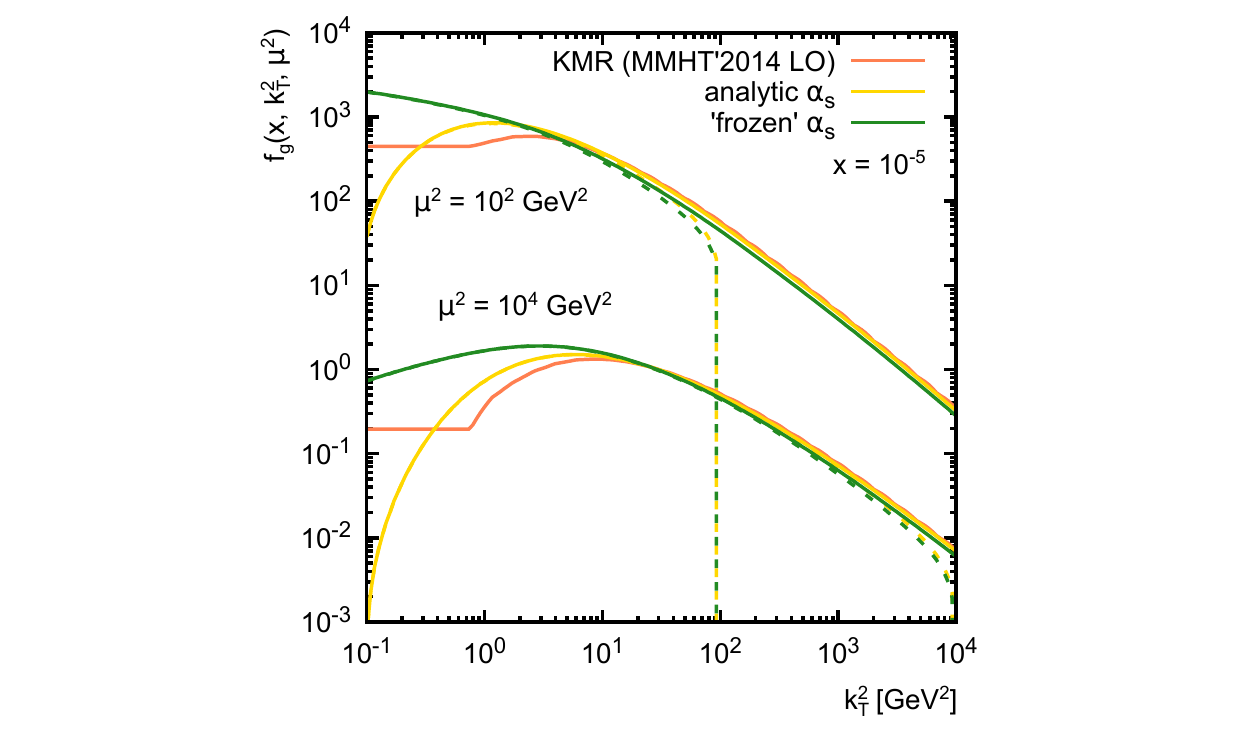}
\includegraphics[width=7.9cm]{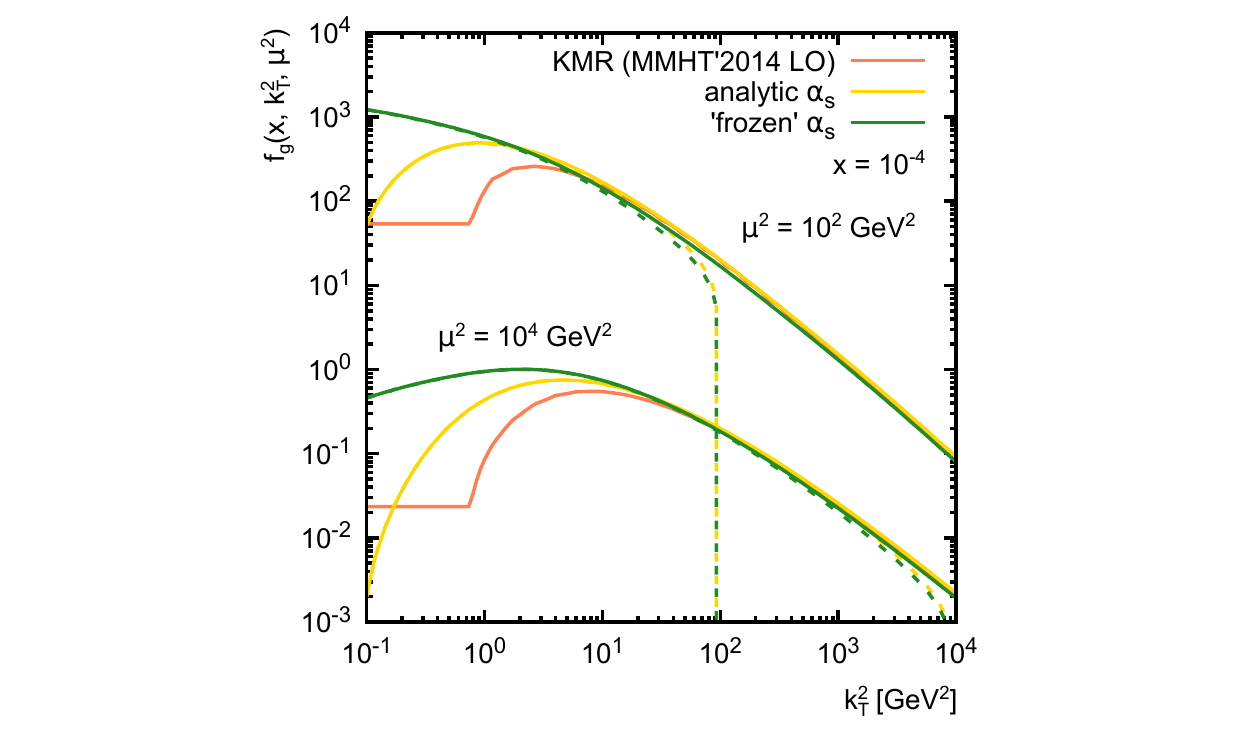}
\includegraphics[width=7.9cm]{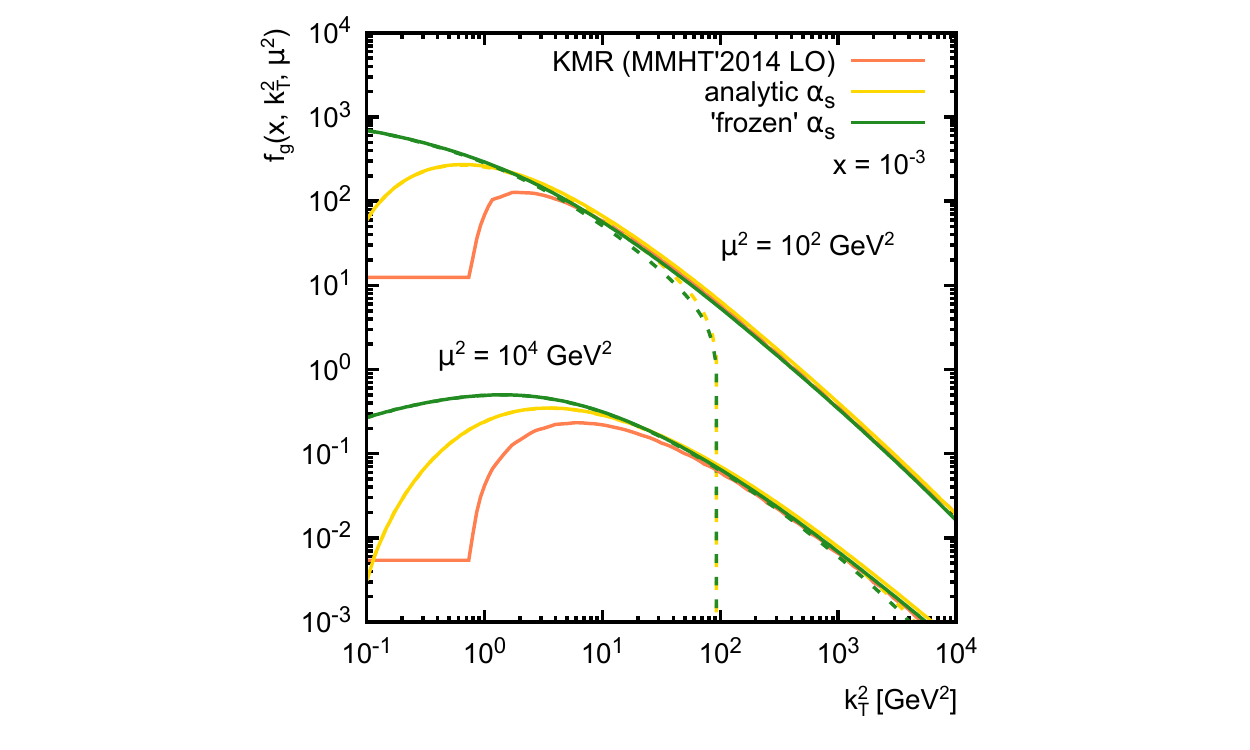}
\includegraphics[width=7.9cm]{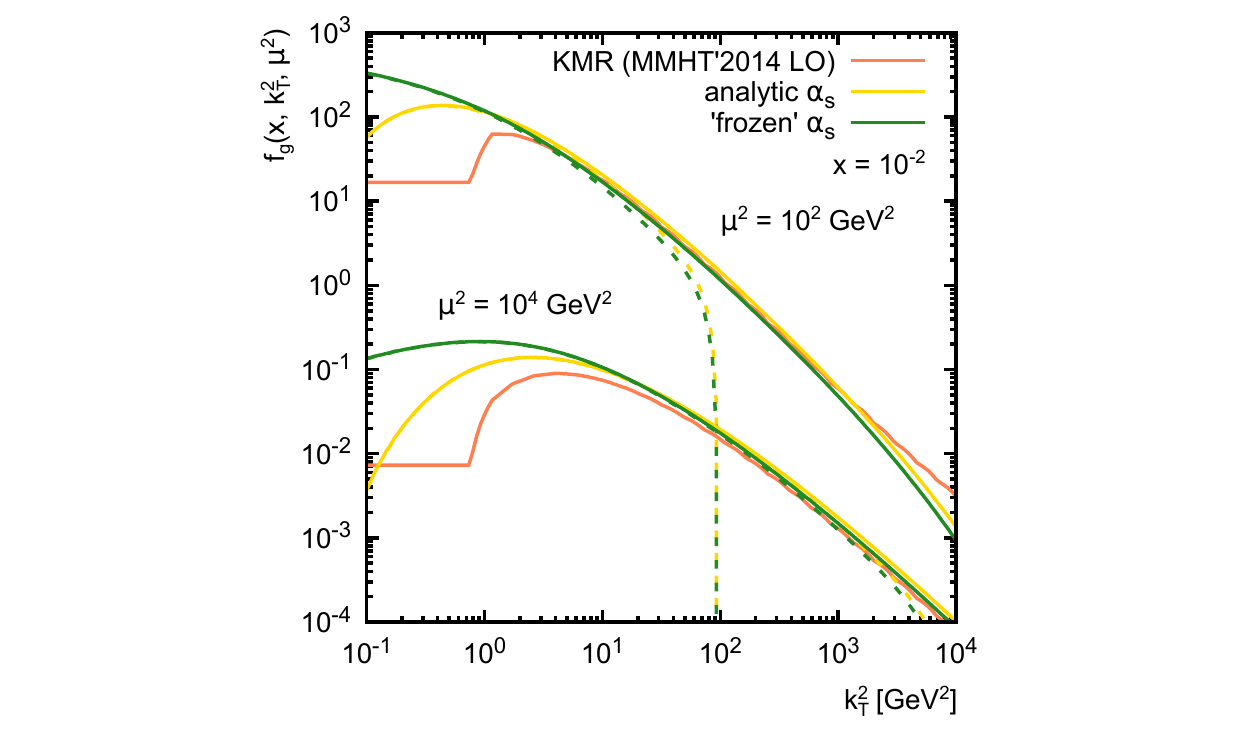}
\caption{The TMD gluon densities in a proton calculated as a function of the gluon
transverse momentum $k_T^2$ at different values of
longitudinal momentum fraction $x$ and hard scale $\mu^2$.
The integral formulation of the KMR approach is used.
The solid green and yellow curves correspond to the
results obtained with "frozen" and analytic QCD coupling constant with
angular ordering condition, while corresponding dashed curves
represent the results obtained with strong ordering condition.
The red curves correspond to the TMD gluon distributions
calculated numerically in the traditional KMR scenario, where the
conventional parton densities from standard MMHT'2014 (LO) set are used as an input.}
\label{fig2}
\end{center}
\end{figure}

\begin{figure}
\begin{center}
\includegraphics[width=7.9cm]{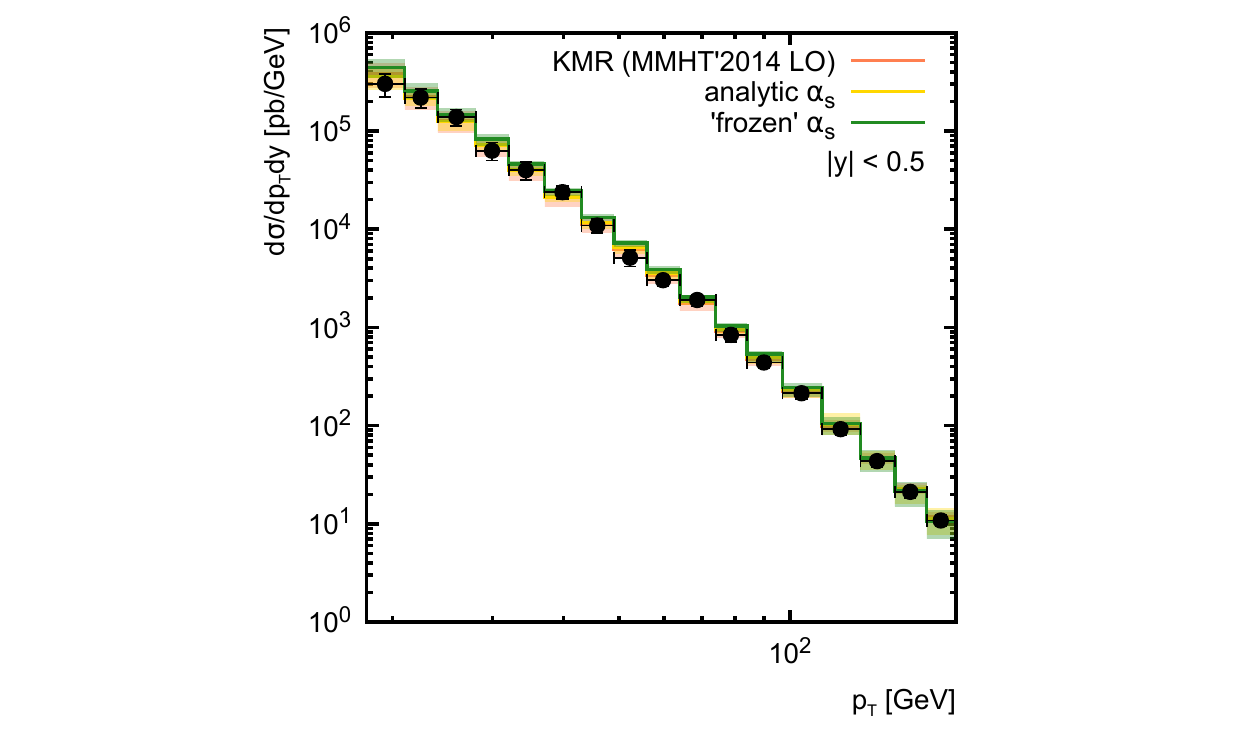}
\includegraphics[width=7.9cm]{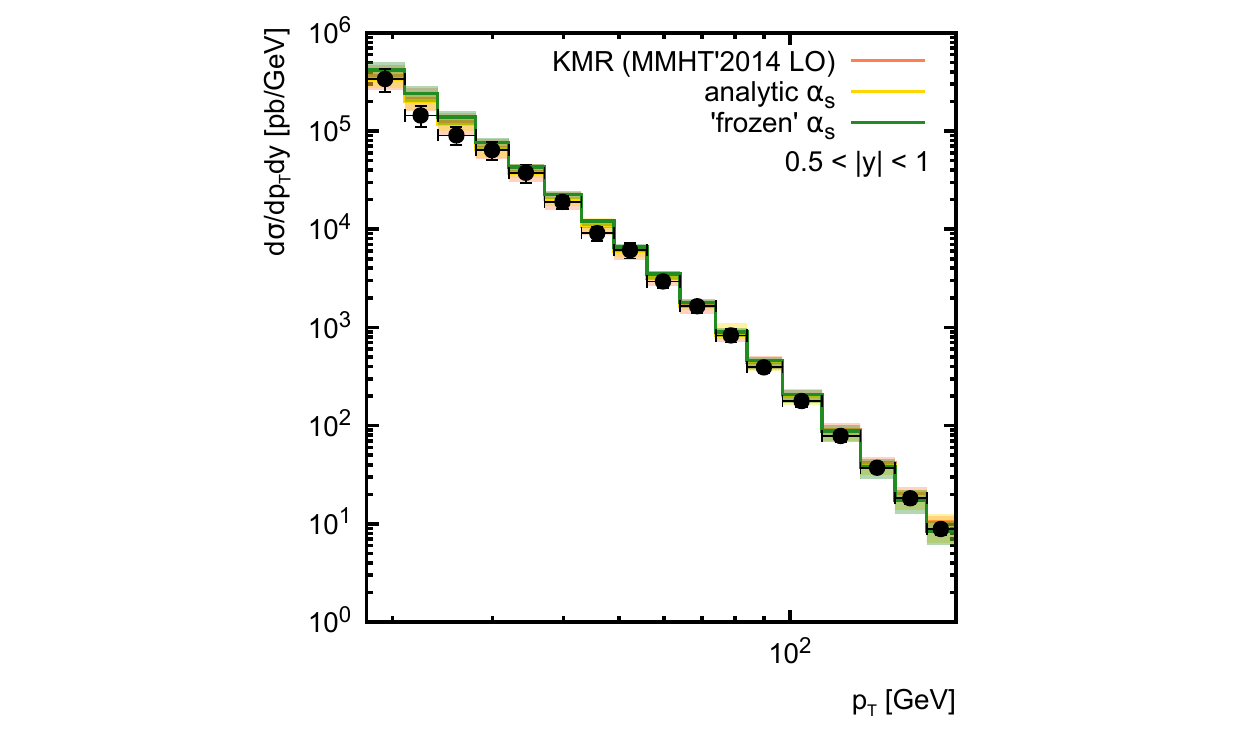}
\includegraphics[width=7.9cm]{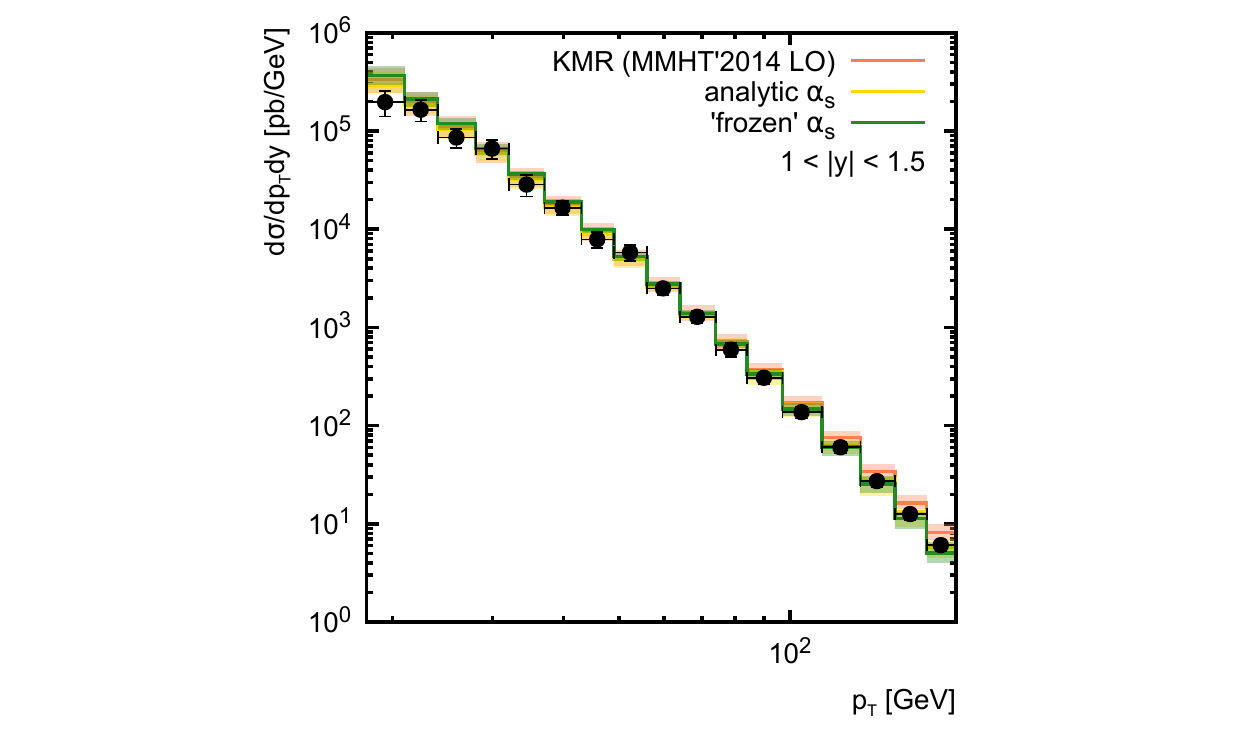}
\includegraphics[width=7.9cm]{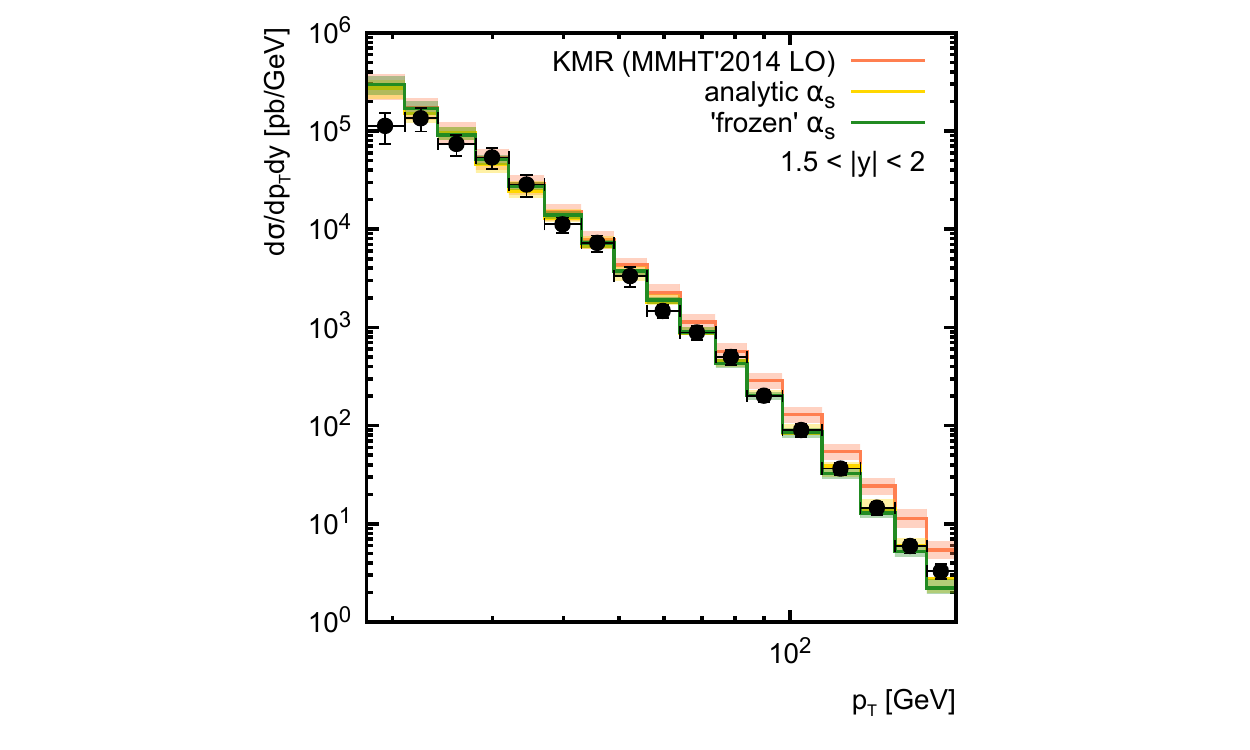}
\includegraphics[width=7.9cm]{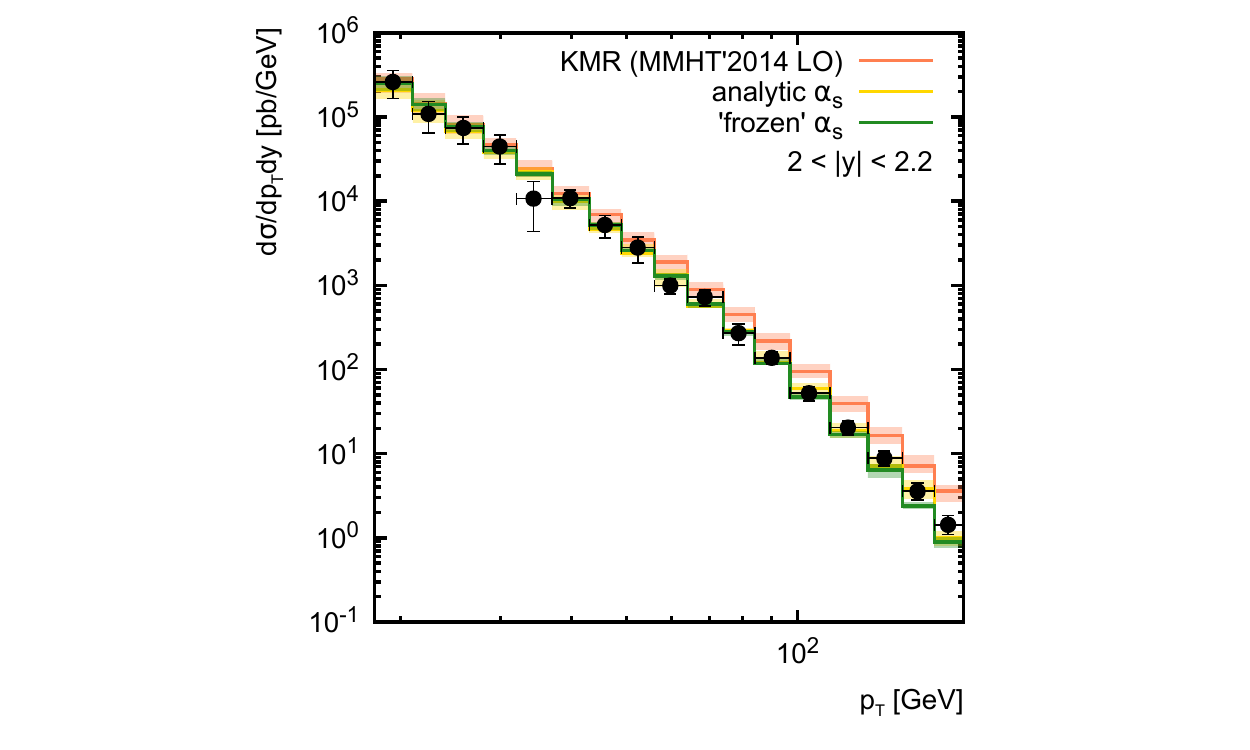}
\caption{The transverse momentum distributions of inclusive $b$-jet production at $\sqrt s = 7$~TeV
as a function of the leading jet transverse momentum in different rapidity
regions. The kinematical cuts are described in the text.
Notation of histograms is the same as in Fig.~1.
The experimental data are from CMS\cite{59}.}
\label{fig3}
\end{center}
\end{figure}

\begin{figure}
\begin{center}
\includegraphics[width=7.9cm]{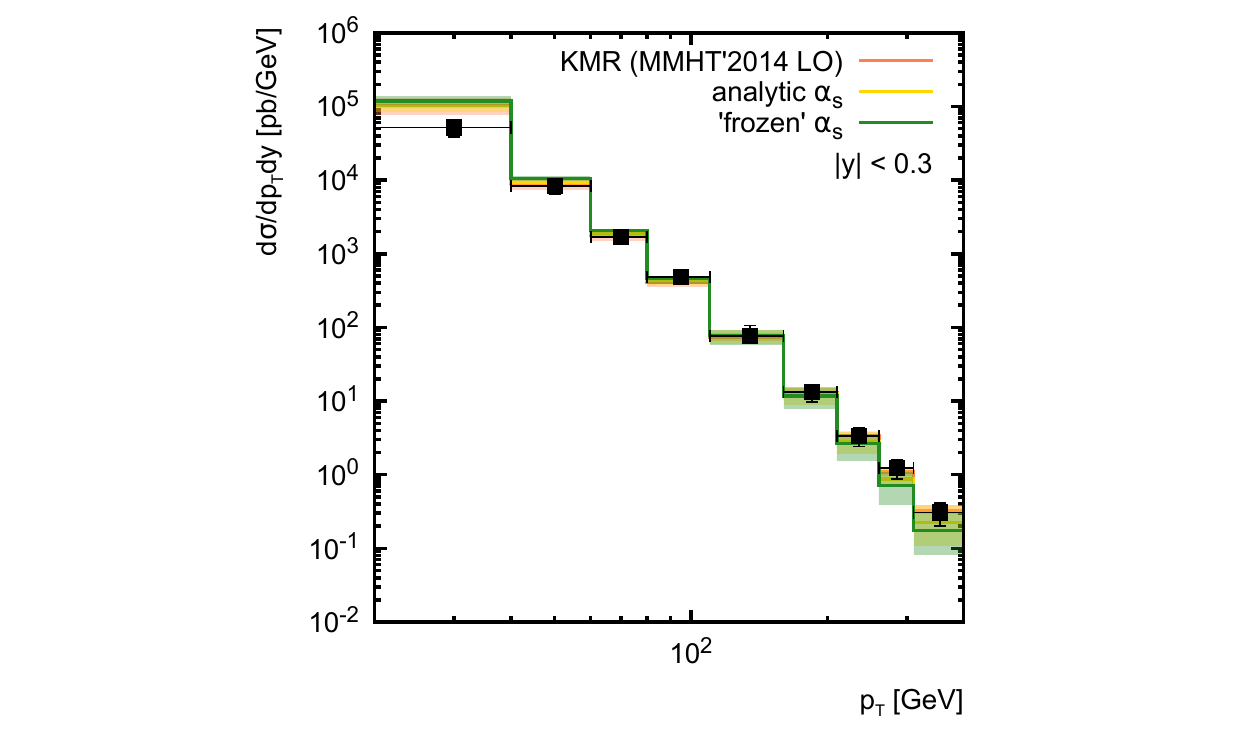}
\includegraphics[width=7.9cm]{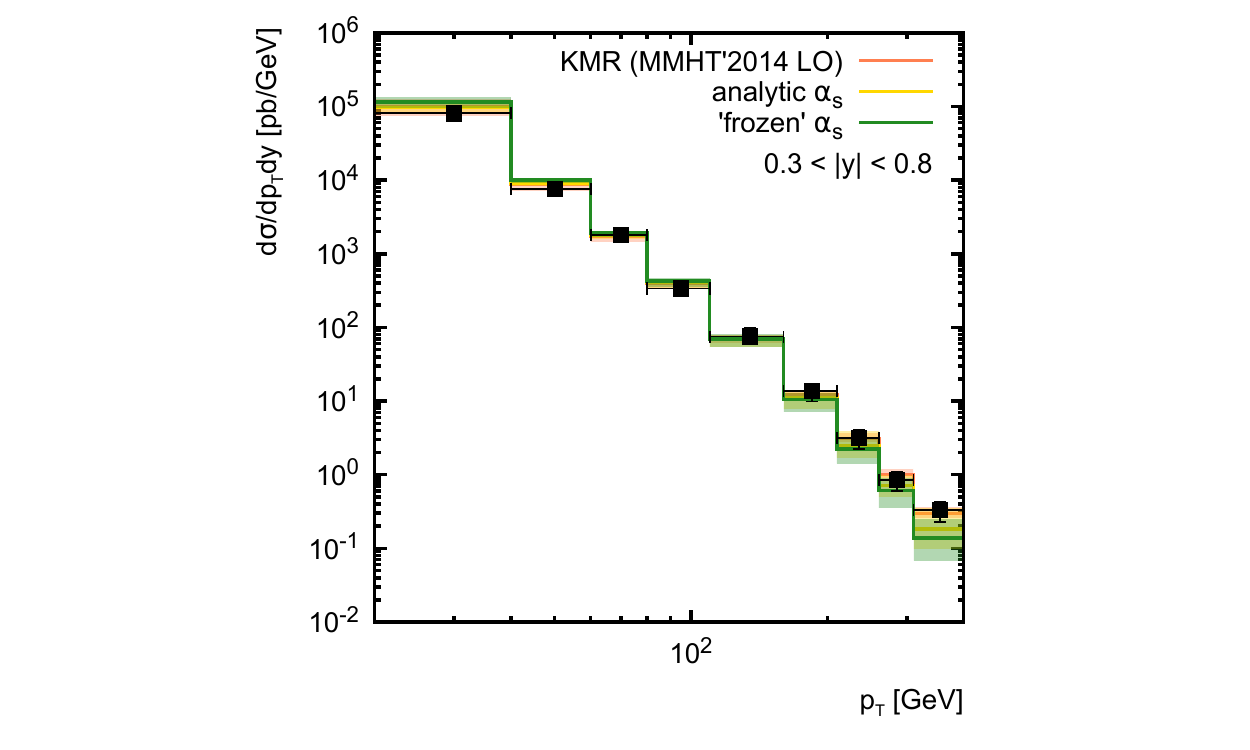}
\includegraphics[width=7.9cm]{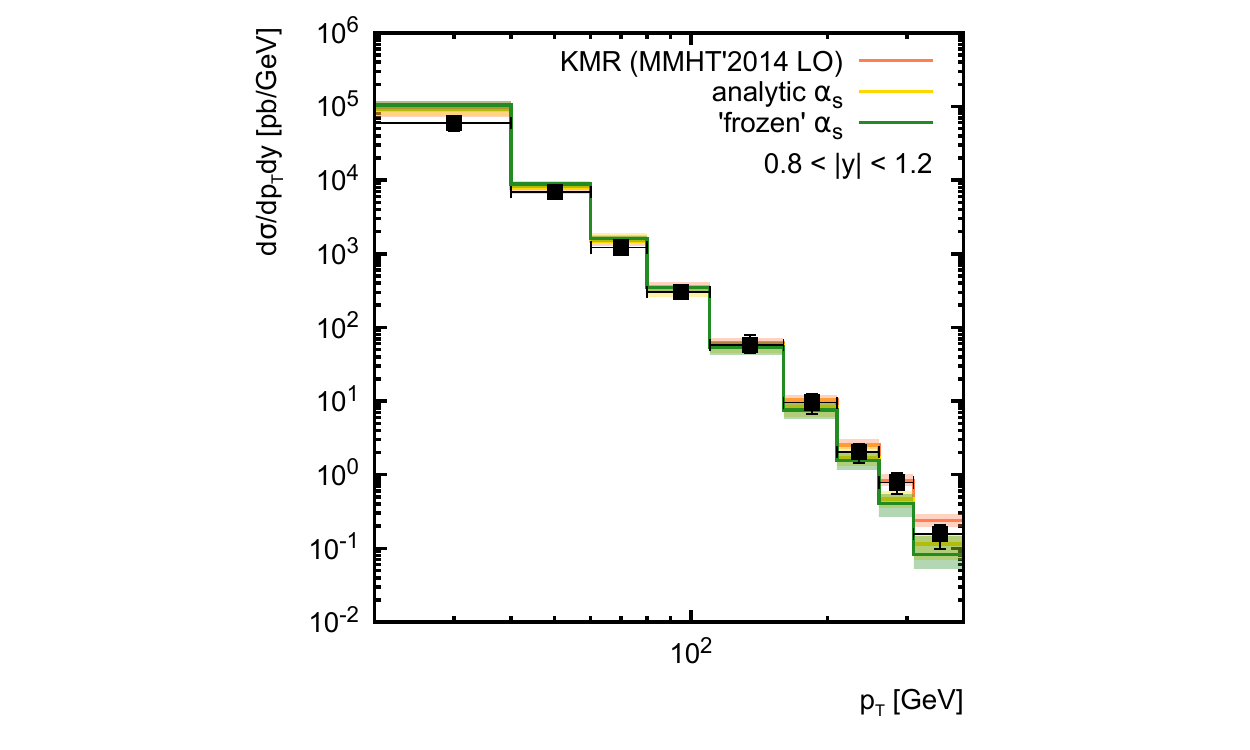}
\includegraphics[width=7.9cm]{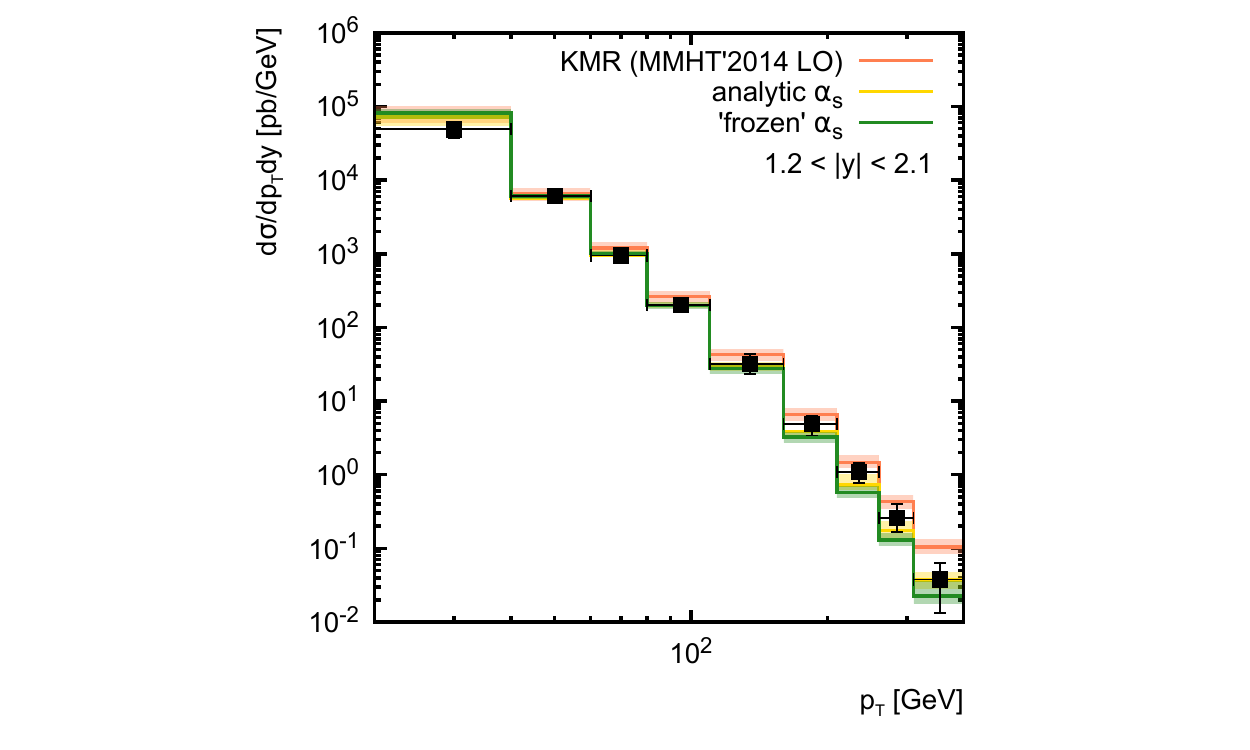}
\includegraphics[width=7.9cm]{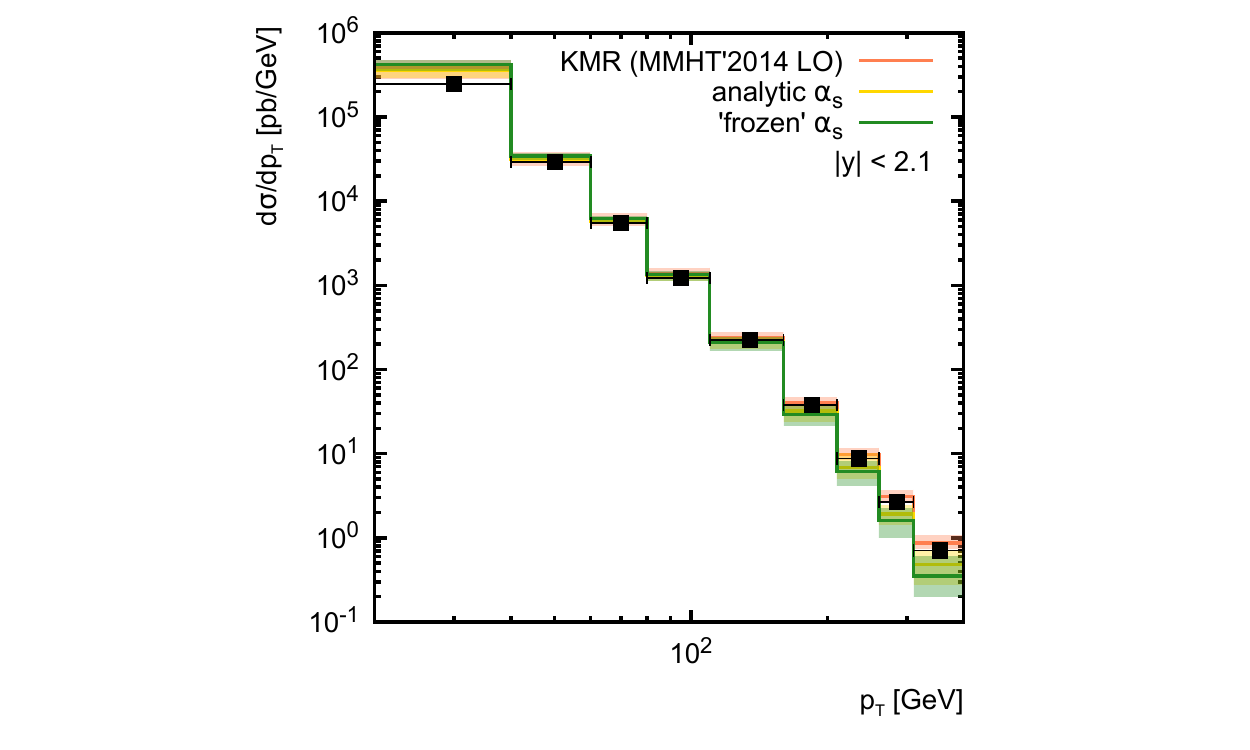}
\caption{The transverse momentum distributions of inclusive $b$-jet production at $\sqrt s = 7$~TeV
as a function of the leading jet transverse momentum in different rapidity
regions. The kinematical cuts are described in the text. Notation of histograms is the same as in Fig.~1.
The experimental data are from ATLAS\cite{60}.}
\label{fig4}
\end{center}
\end{figure}

\begin{figure}
\begin{center}
\includegraphics[width=7.9cm]{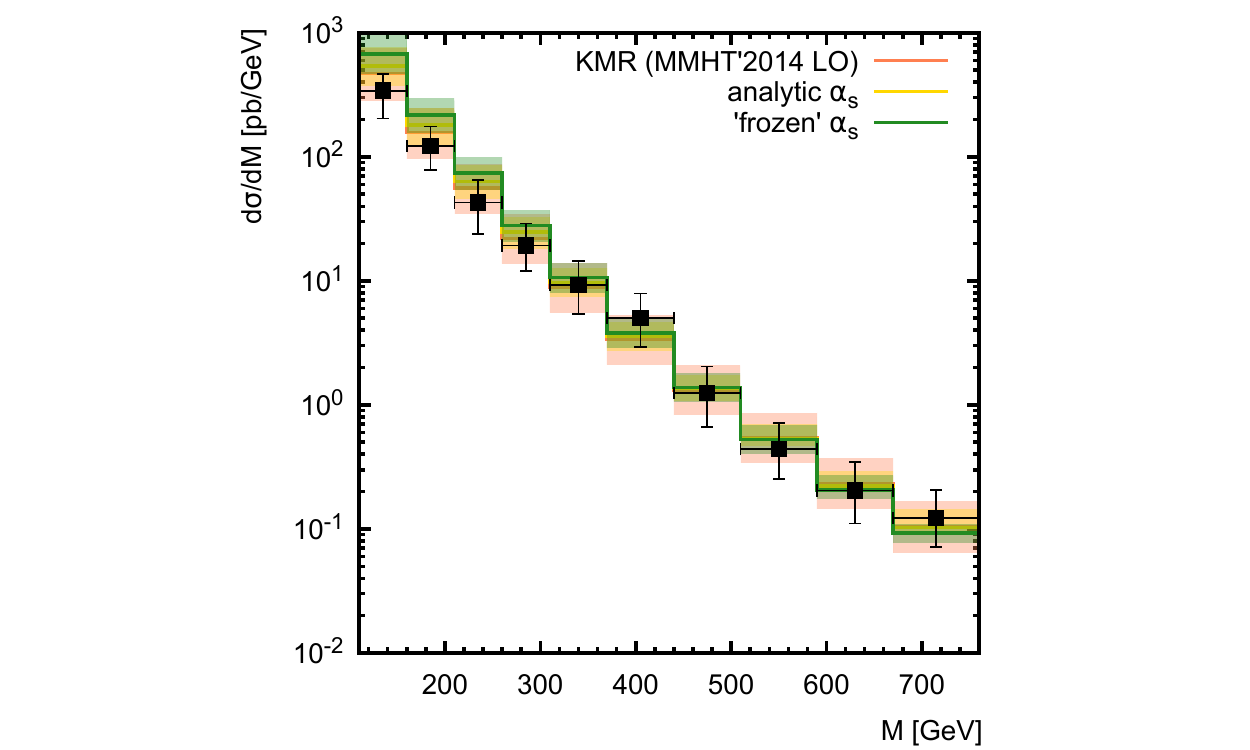}
\includegraphics[width=7.9cm]{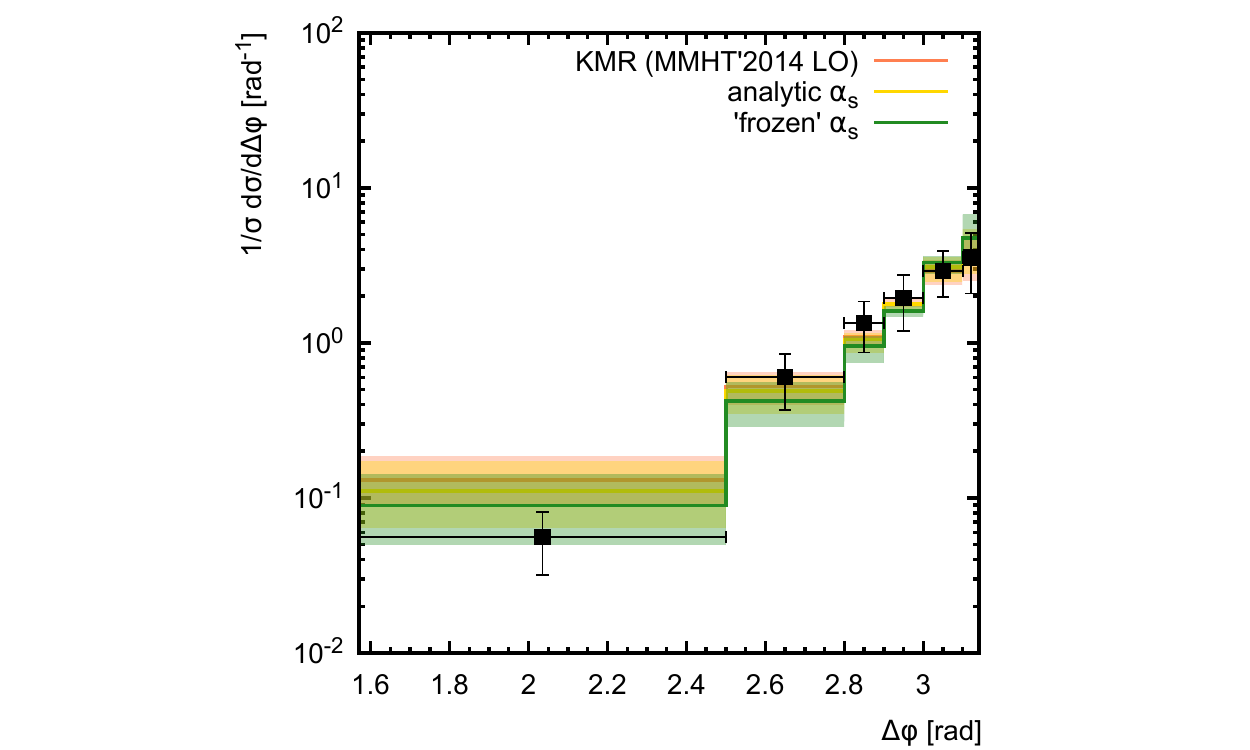}
\includegraphics[width=7.9cm]{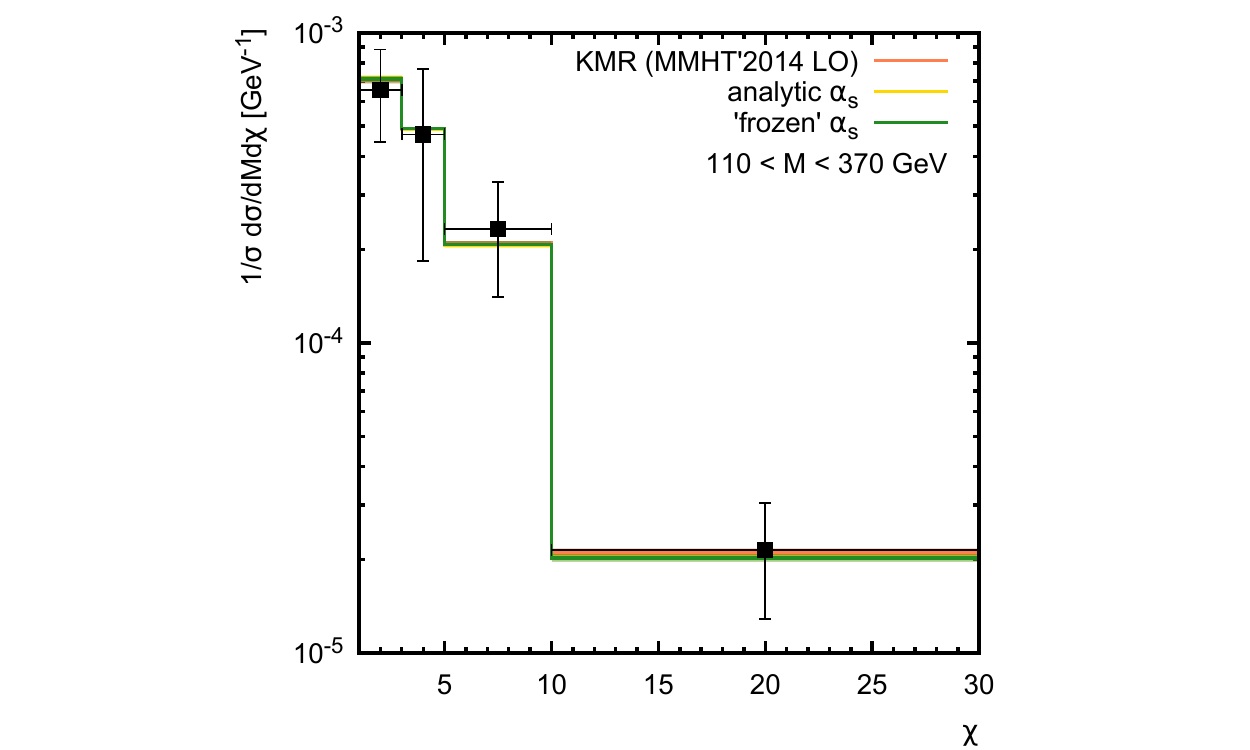}
\includegraphics[width=7.9cm]{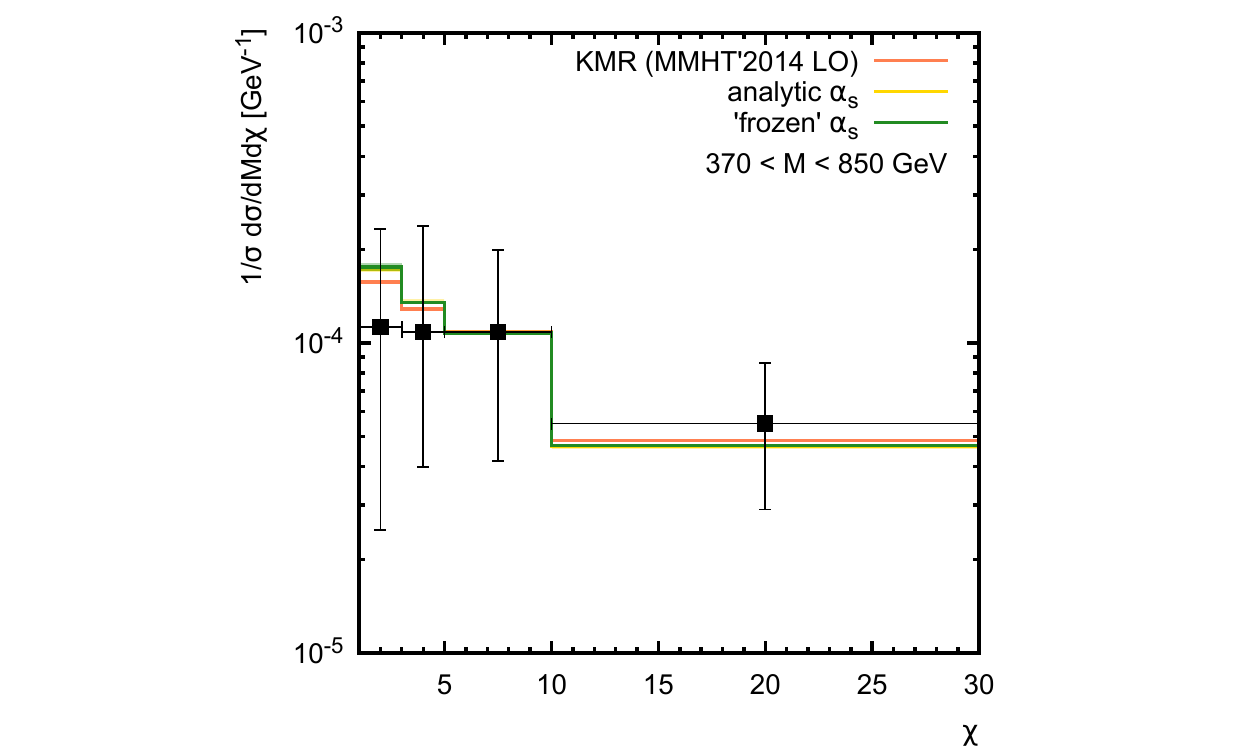}
\caption{The dijet invariant mass $M$, azimuthal angle difference $\Delta \phi$ and
$\chi$ distributions of $b\bar b$-dijet production at $\sqrt s = 7$~TeV.
The kinematical cuts are described in the text.
The experimental data are from ATLAS\cite{60}.}
\label{fig5}
\end{center}
\end{figure}

\begin{figure}
\begin{center}
\includegraphics[width=7.9cm]{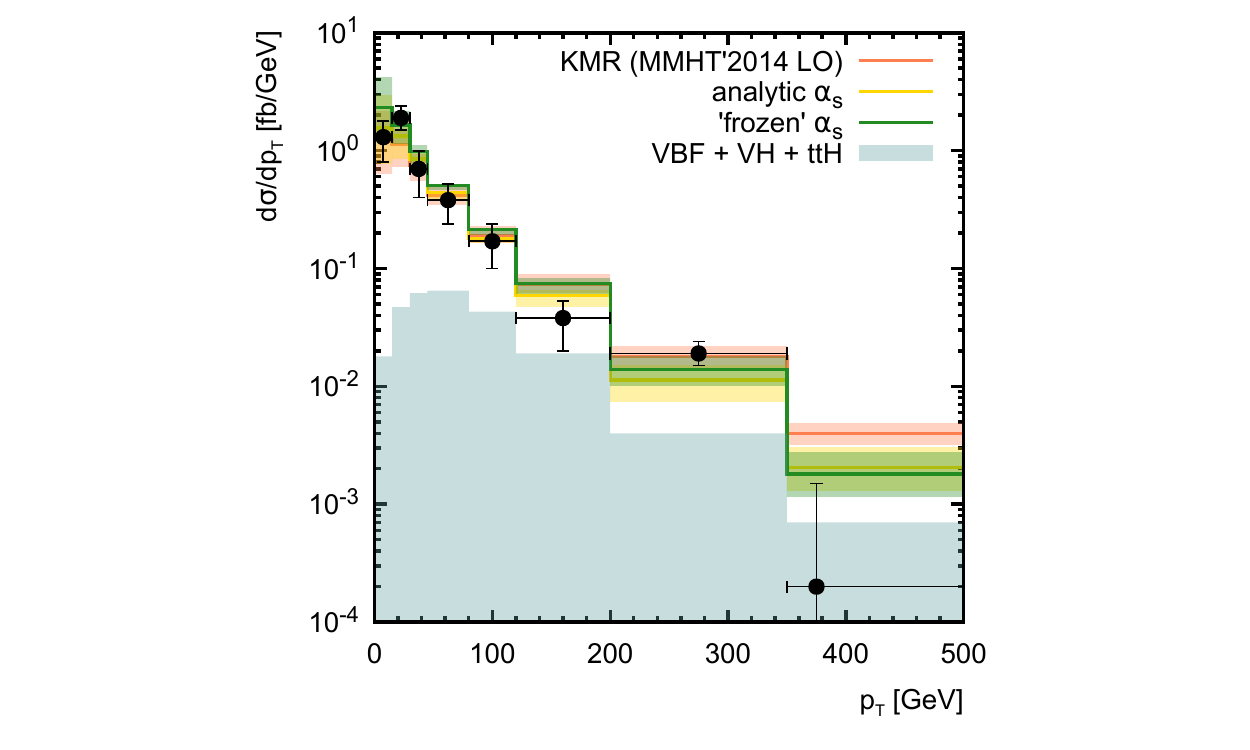}
\includegraphics[width=7.9cm]{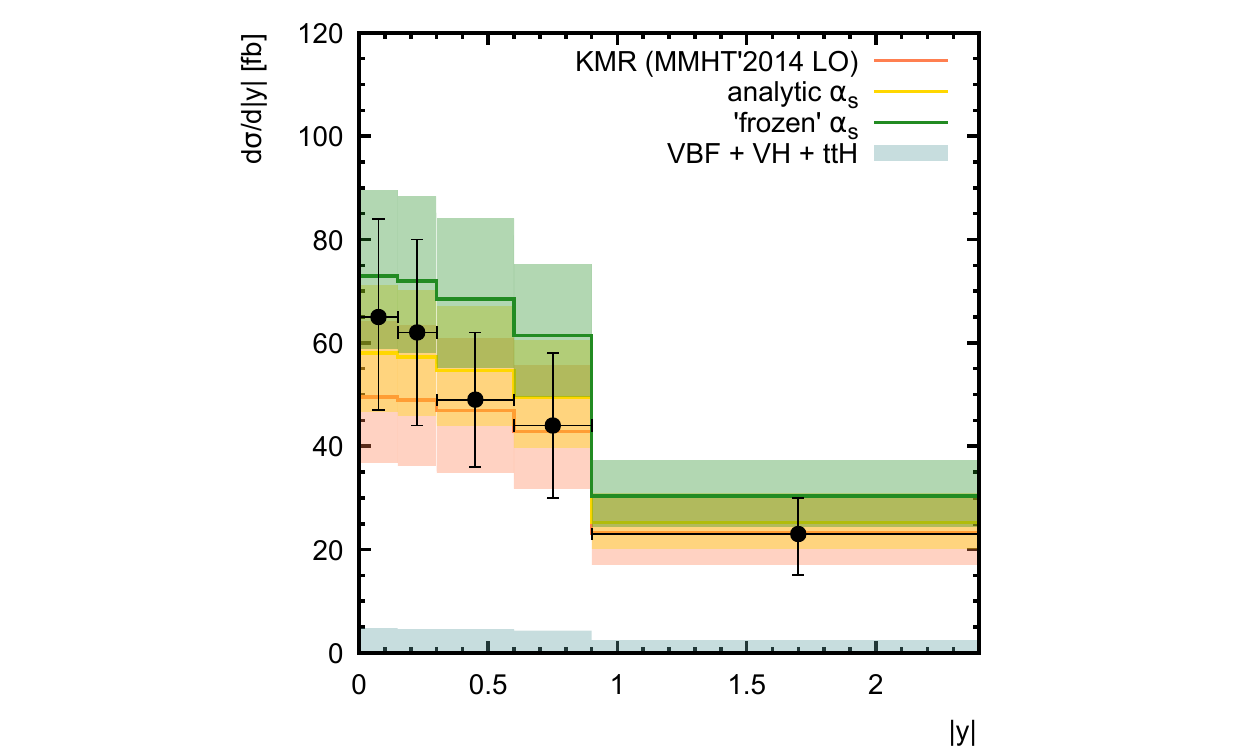}
\includegraphics[width=7.9cm]{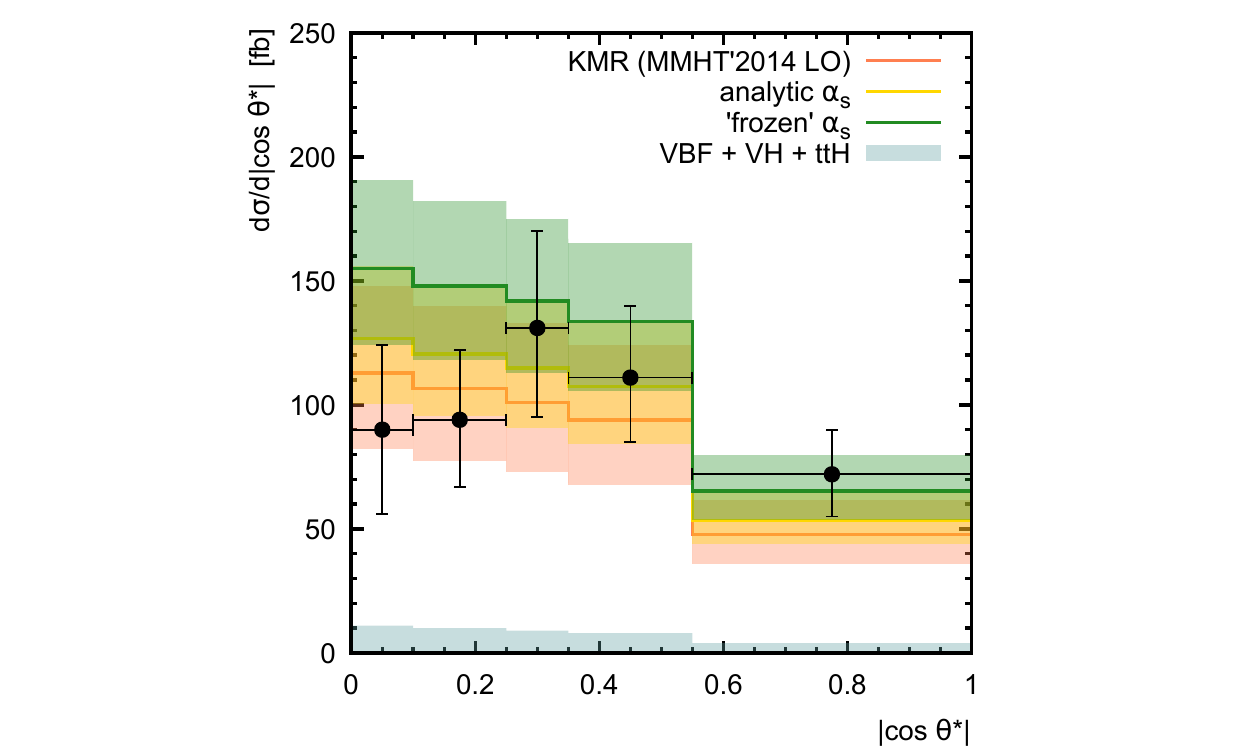}
\caption{The differential cross sections of inclusive Higgs boson production (in the diphoton
decay mode) at $\sqrt s = 13$~TeV as functions of diphoton pair transverse momentum $p_T^{\gamma \gamma}$,
rapidity $y^{\gamma \gamma}$ and photon
helicity angle $\cos \theta^*$ (in the Collins-Soper frame).
The experimental data are from CMS\cite{67}.}
\label{fig6}
\end{center}
\end{figure}

\begin{figure}
\begin{center}
\includegraphics[width=7.9cm]{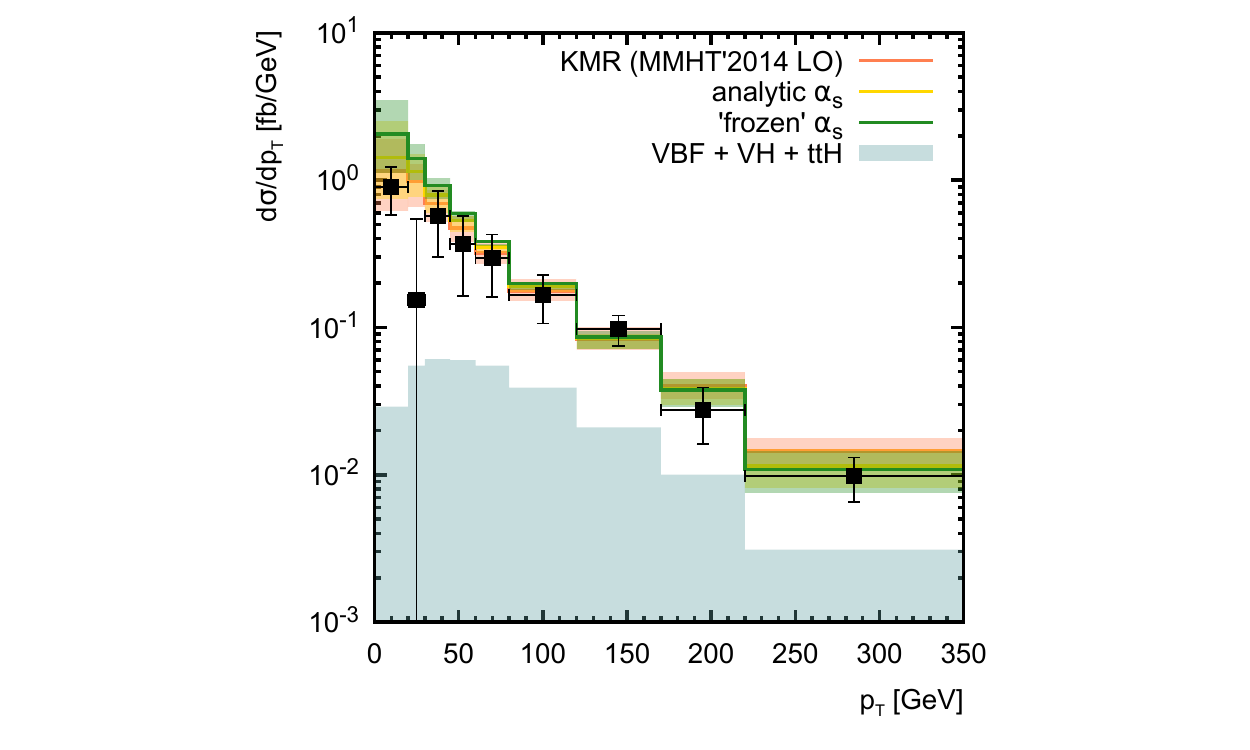}
\includegraphics[width=7.9cm]{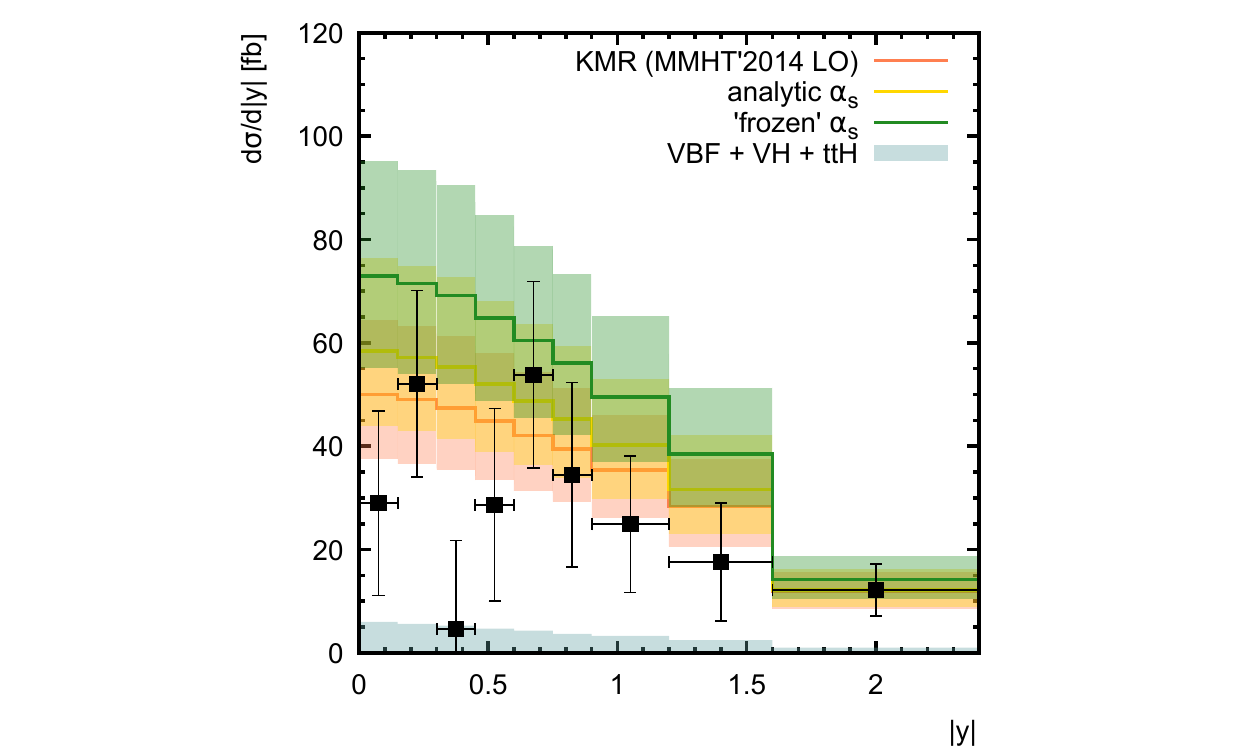}
\includegraphics[width=7.9cm]{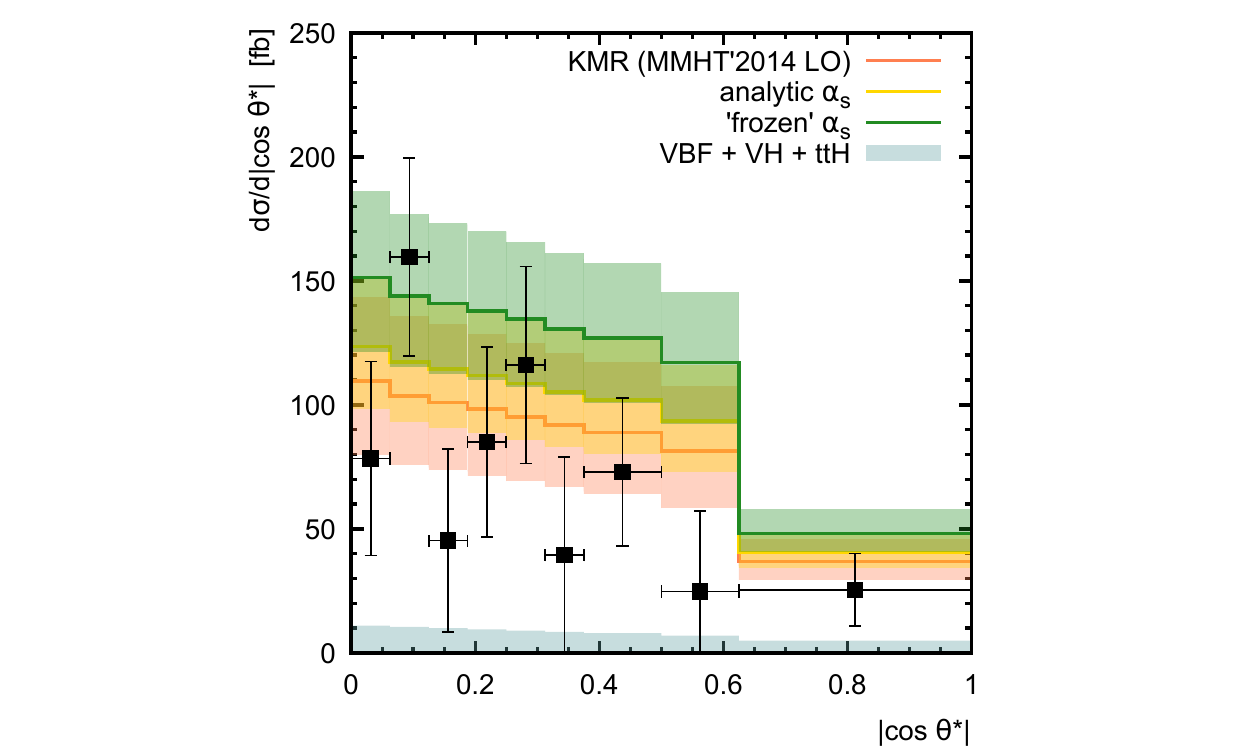}
\caption{The differential cross sections of inclusive Higgs boson production (in the diphoton
decay mode) at $\sqrt s = 13$~TeV as functions of diphoton pair transverse momentum $p_T^{\gamma \gamma}$,
rapidity $y^{\gamma \gamma}$ and photon
helicity angle $\cos \theta^*$ (in the Collins-Soper frame).
The experimental data are from ATLAS\cite{68}.}
\label{fig7}
\end{center}
\end{figure}

\begin{figure}
\begin{center}
\includegraphics[width=15cm]{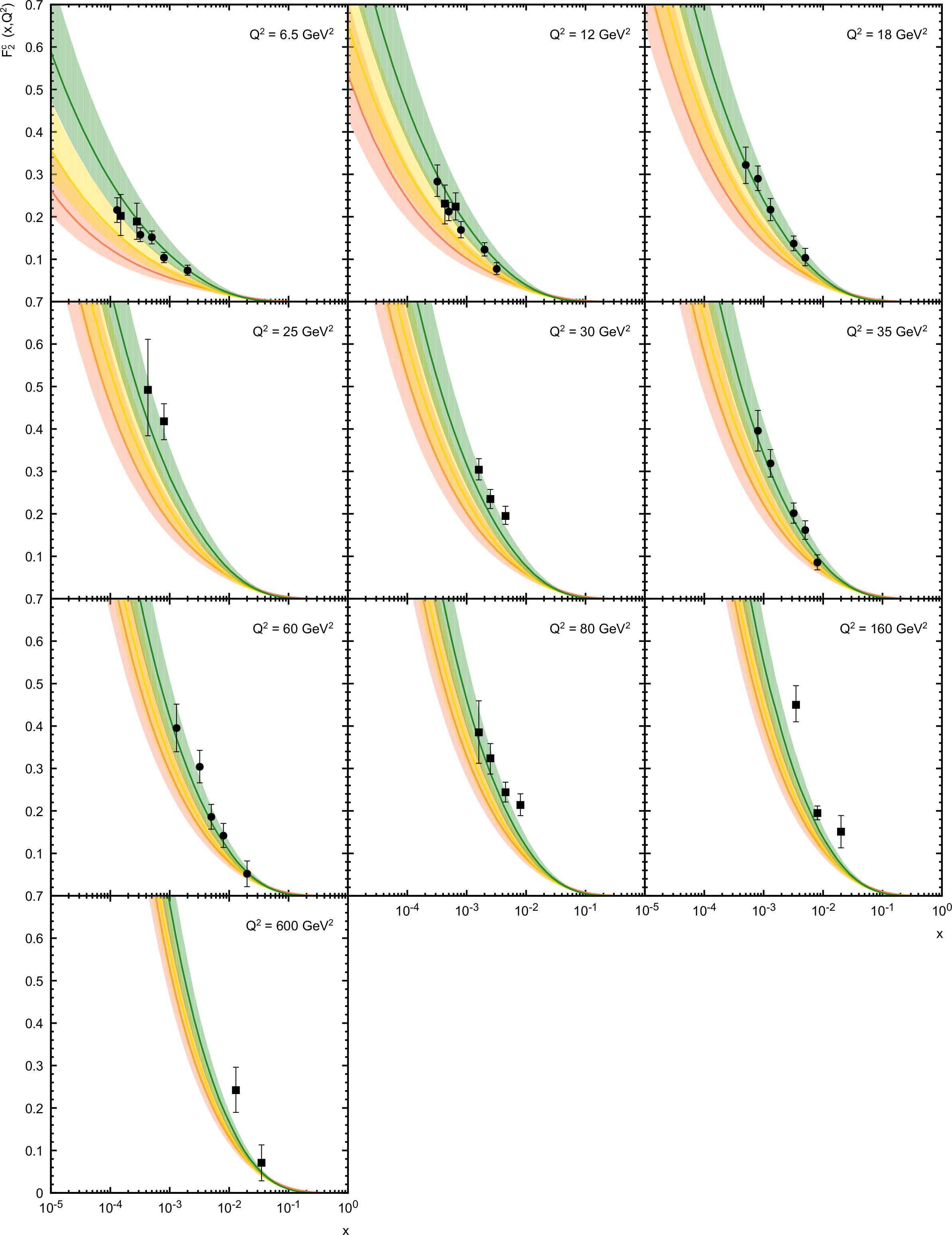}
\caption{The charm contribution to the proton structure function $F_2(x, Q^2)$
as a function of $x$ calculated at different $Q^2$. Notation of curves is the same as in Fig.~1.
The experimental data are from ZEUS\cite{69} and H1\cite{70}.}
\label{fig8}
\end{center}
\end{figure}

\begin{figure}
\begin{center}
\includegraphics[width=15cm]{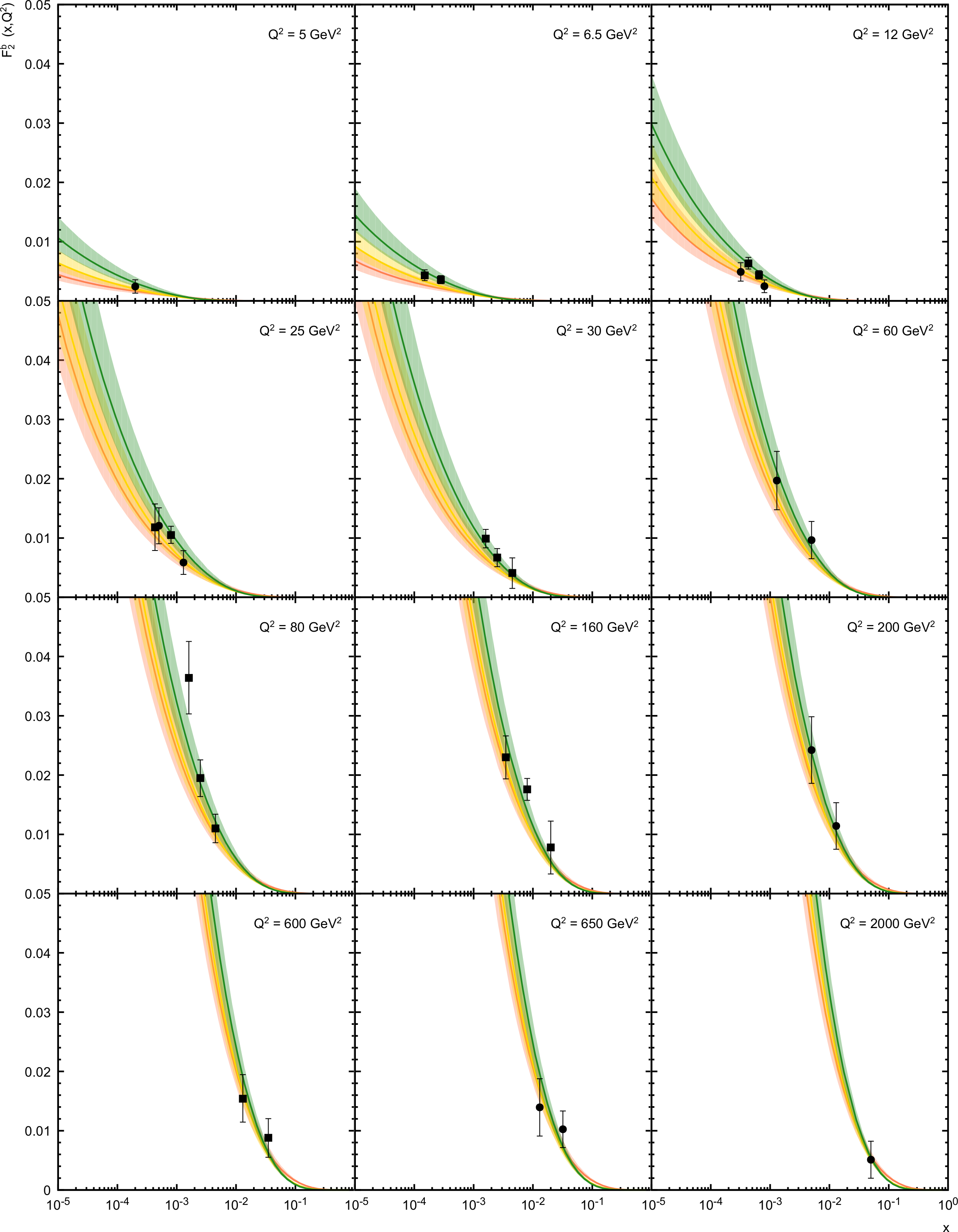}
\caption{The beauty contribution to the proton structure function $F_2(x, Q^2)$
as a function of $x$ calculated at different $Q^2$. Notation of curves is the same as in Fig.~1.
The experimental data are from ZEUS\cite{69} and H1\cite{71}.}
\label{fig9}
\end{center}
\end{figure}

\end{document}